\documentclass[aps, prl, reprint, longbibliography, superscriptaddress]{revtex4-1}

\usepackage[version=3]{mhchem} 
\usepackage[T1]{fontenc}       
\usepackage{color}
\usepackage{multirow}

\usepackage{pdfpages}
\usepackage{pgffor}

\usepackage{amssymb} 
\usepackage{amsmath}
\usepackage{mathrsfs} 
\usepackage{physics} 
\usepackage[per-mode=symbol]{siunitx} 
\usepackage{graphicx} 
\usepackage{hyperref} 
\newcommand*{\eps}[0]{\epsilon}

\newcommand*{\qe}[0]{{\sc Quantum ESPRESSO}}

\newcommand*{\mrm}[1]{{\mathrm{#1}}}

\makeatletter
\AtBeginDocument{\let\LS@rot\@undefined}
\makeatother

\begin{document}

\author{Peter Scherpelz}
\email{pscherpelz@uchicago.edu}
\affiliation{Institute for Molecular Engineering, The University of Chicago,
Chicago, IL, USA}
\author{Giulia Galli}
\affiliation{Institute for Molecular Engineering, The University of Chicago,
Chicago, IL, USA}
\affiliation{Materials Science Division, Argonne National Laboratory, Argonne,
IL, USA}

\title{Optimizing surface defects for atomic-scale electronics: Si
dangling bonds}

\date{\today}

\begin{abstract}
    Surface defects created and probed with scanning tunneling microscopes are a
    promising platform for atomic-scale electronics and 
    quantum information technology applications. Using first-principles
    calculations we demonstrate how to engineer dangling bond (DB) defects on
    hydrogenated Si(100) surfaces, which give rise to isolated impurity states
    that can be used in atomic-scale devices.  In particular we show
    that sample thickness and biaxial strain can serve as control parameters to
    design the electronic properties of DB defects. While in thick Si samples
    the neutral DB state is resonant with bulk valence bands, ultrathin samples
    (1--2 nm) lead to an isolated impurity state in the gap; similar behavior is
    seen for DB pairs and DB wires. Strain further isolates the DB from the
    valence band, with the response to strain heavily dependent on sample
    thickness. These findings suggest new methods for tuning the properties of
    defects on surfaces for electronic and quantum information applications.
    Finally, we present a consistent and unifying interpretation of many results
    presented in the literature for DB defects on hydrogenated silicon surfaces,
    rationalizing apparent discrepancies between different experiments and
    simulations.
\end{abstract}

\maketitle

The ability to engineer semiconducting devices at the atomic scale is key to
achieving further miniaturization of electronics, and to using the
quantum nature of point defects for quantum information applications.  One
promising atomic-scale fabrication method employs scanning tunneling microscopy
(STM) to create and manipulate defects on semiconducting surfaces
\cite{zwanenburg_2013}. For example, dangling bonds (DBs) have been created and
successfully manipulated on hydrogen-terminated Si(100) surfaces by desorbing
individual H atoms from the substrate \cite{lyding_1994}. Ensembles of DBs on
silicon surfaces are now widely used to create atomically-precise systems of
defects \cite{haider_2009, pitters_2011, bellec_2013, schofield_2013, ye_2013,
taucer_2014, kolmer_2014, labidi_2015, kawai_2016, rashidi_2016, hitosugi_1998,
soukiassian_2003, mayne_2006}, leveraging expertise with the fabrication of
silicon devices, including the ability to produce clean and regular
hydrogen-terminated surfaces.

Numerous experiments have demonstrated many attractive properties and potential
applications of DBs on H:Si(100). These defects
interact over next-nearest neighbor distances \cite{haider_2009, pitters_2011,
schofield_2013}, and the charge of individual DBs can be reversibly manipulated,
with given charge states persisting for hours \cite{bellec_2013}. In addition,
these DBs display negative differential resistance, potentially providing a new
component for atomic-scale electronic circuitry \cite{rashidi_2016}. Theoretical
work has suggested that pairs of DBs may be used to create a charge qubit
\cite{livadaru_2010}.   Furthermore, DBs may be assembled into specific patterns
for electronics or quantum simulations \cite{schofield_2013, soukiassian_2003,
ample_2011, kepenekian_2014}, including one-dimensional conducting or
semiconducting wires \cite{hitosugi_1999, naydenov_2013, ye_2013}. They further
serve as a starting configuration for atomically-precise dopant placement
\cite{schofield_2003, ruess_2004, fuechsle_2012, weber_2012, morley_2014}. Hence
tuning and manipulating the properties of DBs, e.g., charge states, may lead to
a promising strategy to build a flexible atomic-scale platform of defects for
electronic and quantum information technology applications.

\begin{figure*}
    \centering
    \includegraphics[scale=0.4]{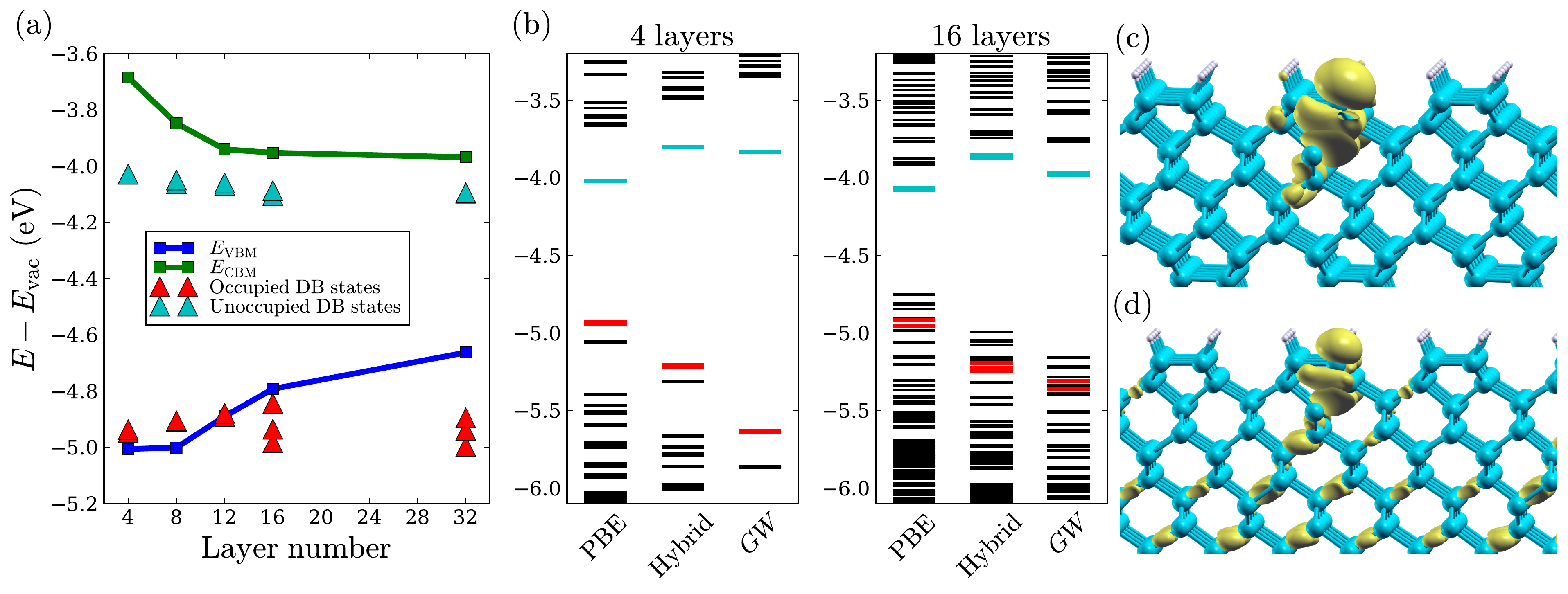}
    \caption{\label{fig:neutraldb}
    Electronic properties of neutrally charged dangling bonds on an H-terminated
    Si(100) slab. \textbf{(a)} Variation of the valence band maximum (VBM),
    conduction band minimum (CBM), and dangling bond (DB) energy levels with
    layer number included in a model Si slab, as obtained using density
    functional theory (DFT) calculations at the PBE level. We estimate that for
    slab thicknesses of \SI{8}{\nano\meter} or greater, the DB position is about
    \SI{0.3}{\electronvolt} below the VBM.
    \textbf{(b)} Energy levels at the $\vb{\Gamma}$ point of the slab Brillouin
    zone, as obtained with DFT using gradient-corrected (PBE) and
    dielectric-dependent hybrid functionals, and many-body perturbation theory
    calculations ($GW$).  Results for 4- and 16-layer slabs are displayed. As in
    (a), red (cyan) designates energy levels of occupied (unoccupied) DB states.
    \textbf{(c)} Isosurface of the wavefunction amplitude for an isolated DB
    state in an H-terminated 8-layer Si slab. \textbf{(d)} Isosurface of the
    wavefunction amplitude for a DB state hybridized with a bulk state in an
    H-terminated 16-layer Si slab. See SM for details. 
    }
\end{figure*}

In this work, using the results of first-principles calculations we propose
ways to realize DB defect states on H:Si(100) with energies within the
electronic gap of bulk Si; in particular we show how to tune sample thickness
and strain to obtain desired energy and charge states. While doing so, we also
address existing controversies present in the literature on the properties of DB
states on Si surfaces. We present a consistent interpretation
of previous results, and use advanced methods to ensure our findings are robust.
We first show how the thickness of the Si sample can be manipulated to alter the
electronic properties of the neutral Si DB state, and we demonstrate that a
stable positively-charged DB state is accessible only in thin
(\SI{1.2}{\nano\meter}) Si samples. We also present similar effects for multiple
DB systems. We then turn our attention to the effect of biaxial strain, showing
that the electronic response of defects to strain depends significantly on the
slab thickness.  We propose that by combining thickness and strain, one may
engineer the properties of neutral DB defects for use in atomic-scale
electronics.

\textit{Methods:} We carried out density functional theory (DFT)
\cite{hohenberg_1964, kohn_1965} calculations with plane-wave basis sets and
norm-conserving pseudopotentials \cite{troullier_1991, ceresoli_2016} using
the \qe{} package \cite{giannozzi_2009, qe_web}. We modeled Si DBs on an
H-terminated Si(100) slab periodically repeated in two directions and
having a finite number of layers in the third, with vacuum separating periodic
images. All neutral DB calculations were spin-polarized. Atomic geometries were
optimized until forces on the atoms were less than
\SI{0.013}{\electronvolt\per\angstrom}.

We used the gradient-corrected exchange-correlation functional developed by
Perdew, Burke, and Ernzerhof (PBE) \cite{perdew_1996}, as well as hybrid
functionals \cite{kummel_2008}. In particular we adopted a dielectric-dependent
hybrid with the fraction of exact exchange $\alpha = 0.085 \approx
1/\eps_\infty^\mrm{Si}$, shown to reproduce accurately the electronic properties
of bulk Si \cite{skone_2014}. We also used the Heyd-Scuseria-Ernzerhof (HSE)
functional \cite{heyd_2003, heyd_2006} to compare with previous work. In
addition, for selected configurations we performed many-body perturbation theory
(MBPT) calculations \cite{hedin_1965, hybertsen_1986, aryasetiawan_1998} at the
$G_0W_0$ level using the WEST software package \cite{govoni_2015,
scherpelz_2016, west_code}. Within $G_0W_0$ the exchange-correlation energy
entering DFT is replaced by an electronic self-energy calculated using the
screened Coulomb interaction and the Green's function.  WEST uses spectral
decomposition techniques \cite{govoni_2015, wilson_2008, wilson_2009,
nguyen_2012, pham_2013} and methods based on density functional perturbation
theory \cite{baroni_2001} to optimize calculations for large systems
\cite{govoni_2015}.

Details of geometries, calculation parameters, convergence
tests, and identification of DB states, can be found in the Supplemental
Material (SM).

%

\textit{Results:} Multiple computational results have been reported in the
literature for the singly-occupied, neutral DB (DB${}^0$) state, relative to the
VBM of Si: \SI{-0.3}{\electronvolt} \cite{wieferink_2010},
\SI{0.013}{\electronvolt} \cite{schofield_2013}, \SI{0.2}{\electronvolt}
\cite{blomquist_2006}, \SI{0.35}{\electronvolt} \cite{livadaru_2010,
haider_2008}, \SI{0.36}{\electronvolt} \cite{ye_2013}, and
\SI{0.42}{\electronvolt} \cite{raza_2007} (other calculations \cite{bellec_2013,
komsa_2013, kawai_2016} only addressed doped systems and/or charged DBs)
\footnote{Only the lowest two values from Refs.\ \cite{wieferink_2010,
schofield_2013} are explicitly spin-polarized; the calculations that cluster at
0.35-0.42 eV are thus unlikely to be physically accurate.}. These results differ
quantitatively and qualitatively: indeed, a defect state located
in energy above the VBM is expected to be well-isolated electronically, whereas
one below may instead hybridize with other electronic states in the material and
hence may not be amenable to manipulation.

In order to rationalize the various literature values for the DB${}^0$, we
calculated its electronic properties for many model slabs, differing by the
number of layers and supercell lattice constant. We found that the choice of
lattice constants in the plane perpendicular to the surface primarily influenced
the degree of dispersion of the DB state (see SM for details).  We observed
instead a much more pronounced dependence of the nature of the DB state on the
thickness of the slab, as shown in Figure \ref{fig:neutraldb}(a).  Its energy
relative to the vacuum energy of the supercell model is roughly constant.
However, the positions of the VBM and conduction band minimum (CBM) vary
significantly with the number of layers in the slab. In general quantum
confinement leads to a larger bandgap whose convergence toward the bulk value
is very slow as a function of the slab layer number
\cite{li_2010, fischetti_2011}. A similar dependence on thickness was recently
found for a bare Si(100)-$p(2{\times}2)$ surface \cite{sagisaka_2017}.

Figure \ref{fig:neutraldb}(a) shows that for a slab 8 layers thick ($\approx
\SI{1.2}{\nano\meter}$) the DB energy is \SI{0.09}{\electronvolt} above the VBM
energy, while for a 16-layer slab ($\approx \SI{2.3}{\nano\meter}$) its
energy is \SI{0.05}{\electronvolt} to \SI{0.19}{\electronvolt} below the VBM
energy. The DB${}^0$ state is well-isolated for 8 or fewer layer slabs, while it
is mixed with bulk Si states for 16 or more layers, as shown in Figure
\ref{fig:neutraldb}(c-d). To verify these findings are robust
with respect to the level of theory used, we carried out additional calculations
using dielectric-dependent hybrid functionals and MBPT.  Figure
\ref{fig:neutraldb}(b) shows that the position of the occupied DB state
relative to the VBM is nearly the same at all levels of theory.  Hybrid and $GW$
calculations significantly correct the bandgap energy found at the PBE level,
but leave the DB state positions relative to the VBM and CBM unchanged.

\begin{figure}
    \centering
    \includegraphics[scale=0.55]{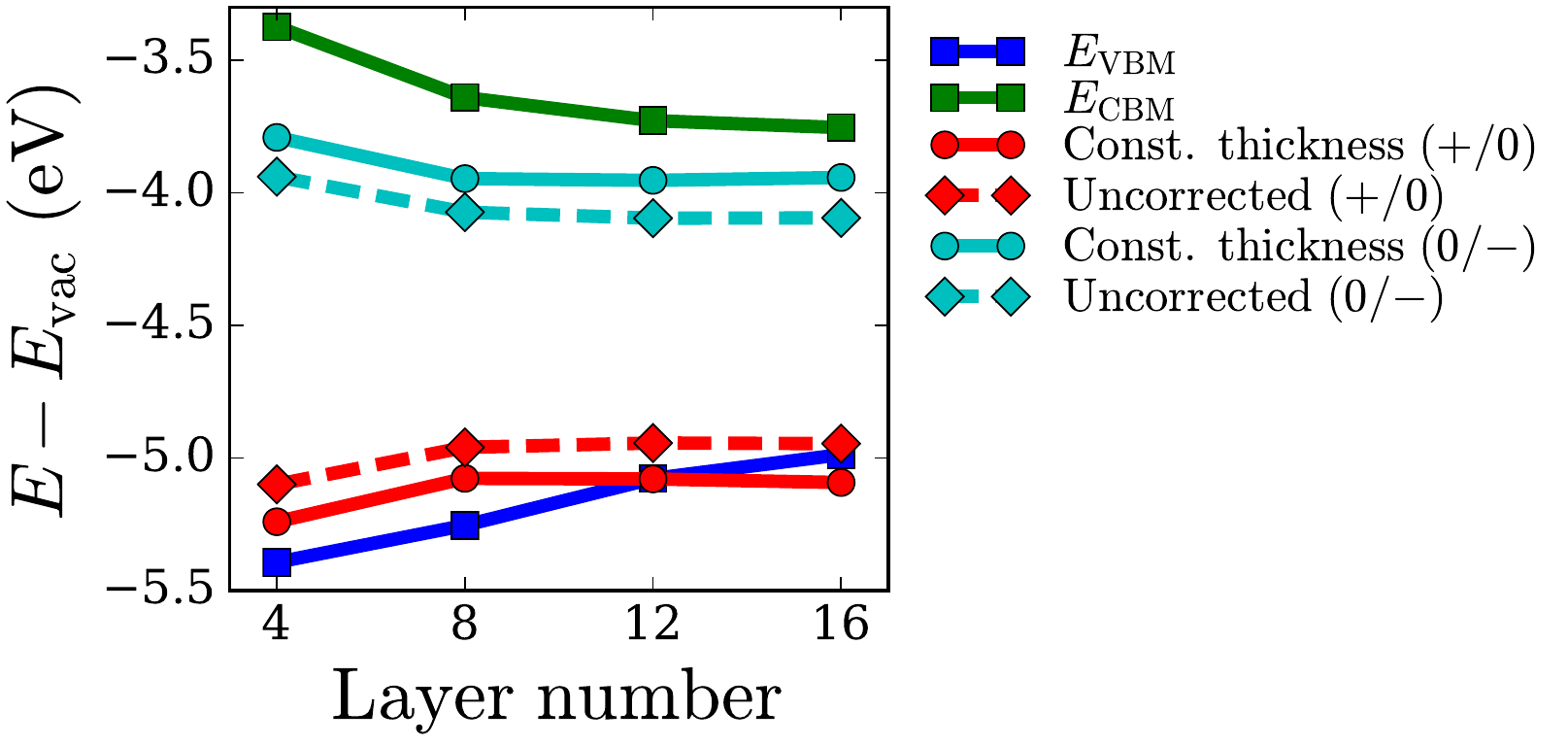}
    \caption{\label{fig:ctls}
    Calculated adiabatic charge transition levels of a DB on an
    H-terminated Si(100) surface, as a function of the sample thickness, using
    the dielectric-dependent hybrid functional. $E_\mrm{VBM}$ and $E_\mrm{CBM}$
    indicate the energies of the valence band minimum and the conduction band
    maximum, respectively. Uncorrected results omit the finite size scaling
    corrections obtained with the method of Ref.~\cite{komsa_2013}.
    }
\end{figure}

Hence we conclude that for Si samples \SI{1.2}{\nano\meter} or thinner, the
neutral singly-occupied DB state falls within the bulk gap, while for samples
\SI{2.3}{\nano\meter} or thicker, it is hybridized with bulk states and resides
below the VBM. We expect the DB${}^0$ state to have a much shorter coherence
time for thickness $ > \SI{2.3}{\nano\meter}$, impacting its behavior in quantum
information applications. However, in other applications, its hybridization with
bulk states of thick slabs may be beneficial by facilitating long-range
interactions between point defects.

We note that these DB properties are different from those of Si DBs at a
Si/\ce{SiO2} interface, for which the energy of the neutral defect is found by
electron paramagnetic resonance to reside in the gap of bulk Si
\cite{lenahan_1998}. This difference is presumably due to the significantly
different environment surrounding DBs in the two systems \cite{raza_2007}. We
also stress that the behavior above is not due to a change in the net
magnetization density (which is not significant as a function of layer
thickness), but rather is due to changes in the energy level of the
singly-occupied DB, which is an isolated defect state for thin slabs but a
resonant defect state for thick slabs.

While understanding the properties of
the neutral DB is important for potential quantum information
applications, charge transition energy levels (CTLs) are the quantities of
interest for scanning tunneling spectroscopy observations \cite{bellec_2013,
ye_2013, taucer_2014, kolmer_2014, labidi_2015, kawai_2016, rashidi_2016}, and
applications for electronics \cite{bellec_2013, rashidi_2016} or charge qubits
\cite{livadaru_2010}. We obtained CTLs by computing total energies of different
charge states in their respective optimized geometries (thus calculating
adiabatic transitions), taking into account corrections for the Coulomb
interaction of periodic images, and alignment of the electrostatic potential
between configurations \cite{freysoldt_2014}. For surfaces, correction
methods used in bulk systems are not applicable due to the large variation in
dielectric constant between the bulk and vacuum. We followed the prescription
suggested in Ref.~\cite{komsa_2013}. Briefly, a sawtooth electric field is used
to compute the $z$-dependent dielectric constant, from which a periodic
electrostatic model of the charged defect is constructed. The electrostatic
energy is then calculated using finite size scaling by extrapolating the energy
of the model computed for cells of increasing sizes.  As we were
interested specifically in the properties of thin Si slabs, we kept the slab
height constant during extrapolation.

\begin{figure}
    \centering
    \includegraphics[scale=0.5]{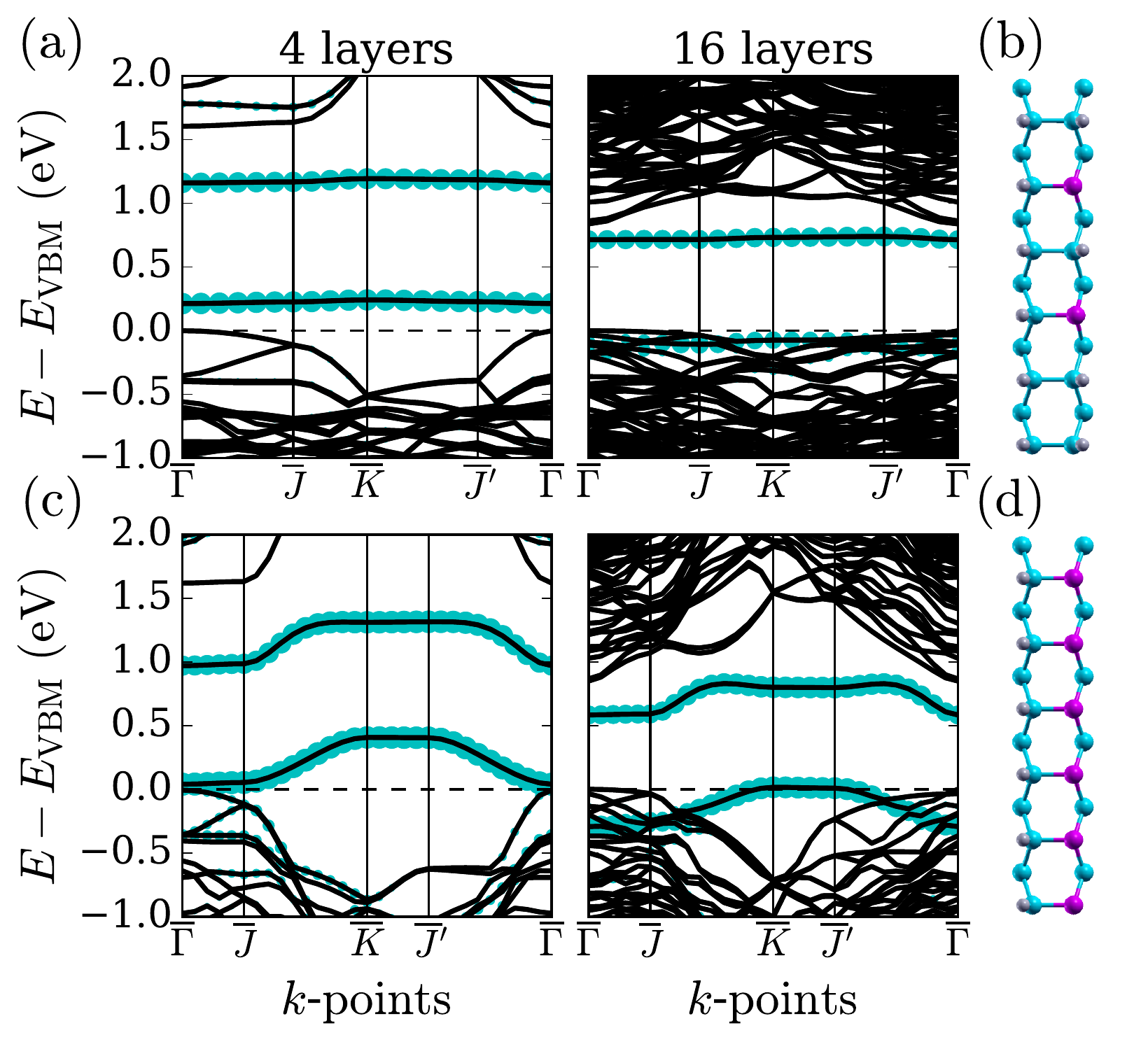}
    \caption{\label{fig:dbsets}
    Electronic properties of multiple-DB systems.  \textbf{(a)} Band structure
    of a neutral DB pair. \textbf{(b)} Model of the DB pair system. Si DBs are
    highlighted. \textbf{(c)} Band structure of the anti-ferromagnetic DB wire.
    Cyan circles in (a) and (c) show the overlap of states with the Si atoms
    containing the DBs.  \textbf{(d)} Model of the DB wire. Si DBs are
    highlighted. The neutral, singly-occupied DBs have alternating spin-up and
	spin-down configurations.
    }
\end{figure}

Calculations at the PBE level of theory were performed to check for convergence,
and consistent results were found, within $\pm\SI{0.1}{\electronvolt}$, when
varying vacuum length of the supercell by a factor of 2 and horizontal
supercell area by a factor of 2.7.  Figure \ref{fig:ctls} shows our results
using the dielectric-dependent hybrid functional. Qualitatively, the PBE results
are similar to those in Figure \ref{fig:ctls}, with a \SI{0.2}{\electronvolt}
decrease in the $({+}/0)$ CTL relative to the VBM (see SM).

Figure \ref{fig:ctls} shows that for thick samples ($>\SI{2.3}{\nano\meter}$),
only the $(0/{-})$ CTL falls within the bulk Si bandgap. This result is
consistent with the experimental observation showing long lifetimes of both the
DB${}^0$ and DB${}^{-}$ charge states, when the DB is appropriately charged by
an STM tip which is then removed \cite{bellec_2013}. No long-lived DB${}^{+}$
state was detected experimentally, even though a $p$-type Si sample was used
\cite{bellec_2013}. Interestingly, Figure \ref{fig:ctls} also shows that very
thin samples may exhibit long-lived DB${}^{+}$ states without requiring any
other perturbations, making all three charge states easily accessible. Such a
system may provide a flexible platform for quantum information technology
applications.

Arrangements of multiple DBs lead to additional possibilities for atomic-scale
electronics. We considered two prototypical multiple-DB systems, and we found
that they exhibit the same properties as a single DB, as a function of
thickness. Figure \ref{fig:dbsets}(a) shows the band structure of a
neutral DB pair \cite{livadaru_2010, pitters_2011, schofield_2013}.
Its geometry (Figure \ref{fig:dbsets}(b)) is an example of that proposed for
charge qubits \cite{livadaru_2010, pitters_2011}.  For a 4-layer slab all DB
states are well-separated, while for a 16-layer slab the
occupied DB states become resonant with the valence band. The presence
of this resonance may lead to stronger interactions between bulk and DB states;
the overall larger gap for a 4-layer system should also improve addressability
of the DB pair using mid-infrared lasers \cite{shaterzadeh-yazdi_2014}.

\begin{figure}
    \centering
    \includegraphics[scale=0.5]{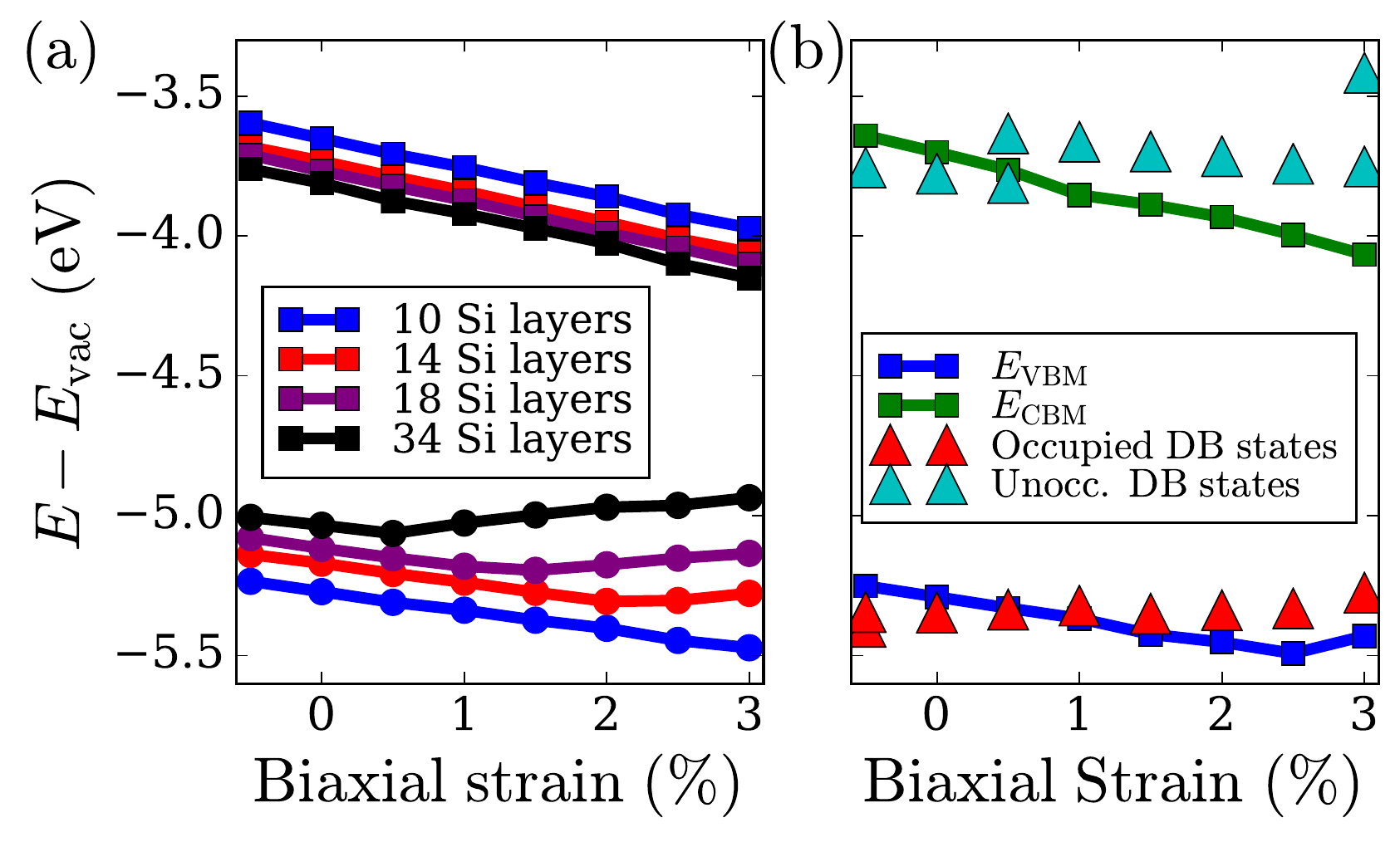}
    \caption{\label{fig:strain}
    Variation of electronic properties as a function of biaxial strain applied
    to the Si slab. \textbf{(a)} Variation of the valence band maximum (VBM,
    circles) and conduction band minimum (CBM, squares) positions vs.\ strain.
    For all values of strain, the bandgap for 34-layer slabs is within 0.1 eV of
    the bulk bandgap.
    \textbf{(b)} Variation of VBM and CBM positions, and the energies of DB
    states, for a 10-layer system.
    }
\end{figure}

Figure \ref{fig:dbsets}(c) shows the band structure of the
anti-ferromagnetic DB wire in Figure \ref{fig:dbsets}(d) \cite{hitosugi_1999,
lee_2008, lee_2009, robles_2012, kepenekian_2013, naydenov_2013, ye_2013,
engelund_2016}.  For a neutral DB wire on a 4-layer slab, both the occupied and
unoccupied one-dimensional (1D) bands lie within the gap, and thus the wire may
conduct either electron or hole states under suitable bias. In contrast, for a
16-layer slab, only the unoccupied 1D band lies within the gap, while the
occupied 1D band is resonant with the bulk Si states. Thus, for 16 layers, hole
conduction would be expected to occur through both bulk and wire states when a
suitable bias is applied, removing the 1D nature of the conductivity.

A recent theoretical study raised the intriguing possibility of using strain to
isolate surface states of a bare Si(100) surface, showing that the VBM is
lowered in energy when biaxial tensile strain is applied in the horizontal
directions \cite{zhou_2013}.  However, this result is not consistent with those
of theoretical and experimental investigations of bulk Si under biaxial tensile
strain, showing that the VBM energy increases and the bulk bandgap decreases
\cite{munguia_2008, richard_2003}. Our results indicate that for standard
(thick) silicon surfaces, strain will not isolate surface states, in contrast to
the conclusions of Ref.~\cite{zhou_2013}. Instead, only 
for thin Si slabs can strain be used to isolate surface states, a result of
the remarkably
different response to strain for thin slabs compared to bulk systems.  Figure
\ref{fig:strain}(a) shows this difference in response: the VBM position as a
function of strain shows a qualitatively different trend for thick 
($\gtrsim\SI{4}{nm}$) slabs, where it increases with strain, and thin (10-layer,
\SI{1.5}{\nano\meter}) slabs, where it decreases with strain to a
thickness-dependent minimum. As a result, for thick slabs the fundamental gap
remains close to that of the bulk, but for thin slabs tensile strain leads to an
increase of the gap relative to that of the bulk.  An unstrained
10-layer slab has a fundamental gap \SI{0.4}{\electronvolt} larger than that of
bulk Si. In contrast, under 3\% biaxial tensile strain its fundamental gap is
\SI{0.8}{\electronvolt} larger than that of the bulk.

Thus, while in thick slabs biaxial tensile strain
would not be expected to aid in isolating DB states, in very thin slabs, the
opposite is true, as shown in Figure \ref{fig:strain}(b). A state which is not
well-isolated in a \SI{1.5}{\nano\meter} unstrained slab becomes isolated for
strains of 1\% or more. For direct comparison with 
Ref.~\cite{zhou_2013}, results
in Figure \ref{fig:strain}
used the HSE hybrid functional;
calculations with PBE showed the same trend as a function of strain
(see SM).
Note that Ref.~\cite{zhou_2013} used a 10-layer slab model,
effectively reporting results applicable only to thin slabs.

In summary, we have proposed how to engineer the properties of DBs on
hydrogenated Si surfaces by varying sample thickness and applied stress.  We
have shown that the single particle energy and wavefunction of DBs on
H:Si(100)-$(2{\times}1)$ may be more readily isolated from those of bulk states
in thin samples ($< \SI{1.2}{\nano\meter}$) than in bulk-like slabs.
Specifically, in thin samples, the neutral DB state
is well above the VBM, and three charge states may be stabilized.  In thick ($>
\SI{2.3}{\nano\meter}$) samples, the neutral DB state is instead hybridized with
bulk states, and the positively charged DB is not stable; for bulk samples the
neutral DB state is about \SI{0.3}{\electronvolt} below the VBM. We verified
that our results are robust with respect to the level of first-principles theory
used, including semi-local and hybrid functionals and many-body perturbation
theory.

Dangling bond pairs and wires showed the same response to sample
thickness. Notably, thin samples allow hole conduction along isolated DB
wires, whereas in thick samples conduction would also occur through the bulk. 
Additionally, we found that in thin samples
biaxial tensile strain will further isolate the 
DB energy from that of the VBM. However, strain is not helpful in
isolating states in thick Si samples.

We emphasize the importance of carrying out accurate calculations, numerically
converged and at a high level of theory, in order to determine the 
properties of isolated DBs.  Although numerous experimental STM studies have
been performed, both tip-induced band bending and non-equilibrium charging did
not allow for a clear extrapolation of results to isolated DB configurations
\cite{bellec_2013, schofield_2013, ye_2013, taucer_2014, kolmer_2014,
labidi_2015, kawai_2016, rashidi_2016}.  Furthermore, recent work has shown that
nearby dopants can affect the behavior of DBs \cite{labidi_2015}, which further
complicates the interpretation of experimental findings.  Our study of
electronic properties as a function of film thickness was able to reconcile
apparent discrepancies found in published results, which were reported for
different numbers of layers in the slabs and sometimes interpreted as
representative of bulk samples (we estimate that 60 layers are necessary for
calculations to be representative of thick, bulk-like samples).

A question remains on the experimental realization of the thin films proposed
here as promising platforms. Si films as
thin as \SI{3}{\nano\meter} have been reported \cite{liu_2002}; strained
Si-on-insulator samples less than \SI{10}{\nano\meter} thick have also been
fabricated \cite{munguia_2012}. While the 1-2 nm slabs considered here may
require new techniques, their experimental realization appears possible in the
near future.  Finally, we expect the results found here for
DBs may be valid for several other defects when placed in thin Si slabs.
This includes many defects used in quantum information applications, such as
isolated phosphorous \cite{saeedi_2013, pla_2013}, boron \cite{salfi_2016},
bismuth \cite{mohammady_2012, wolfowicz_2013}, or selenium \cite{lonardo_2015,
morse_2016} dopants, as well as patterned surface systems \cite{schofield_2013,
soukiassian_2003, ample_2011, kepenekian_2014}.  Indeed the electronic
properties of the DBs change as a function of thickness due to the change of the
VBM and CBM themselves, not because of a substantial shift of the defect level.
Hence the combination of strain and thickness proposed here to isolate DB
defects and stabilize multiple charge states should be generalizable to 
other types of defects.

\begin{acknowledgments}
    We gratefully acknowledge Hosung Seo, Marco Govoni, David Schuster, Jeff
    Guest, Pamela Pe\~na Martin, John Randall, James Owen, Scott Schmucker,
    Neil Curson and Hannu-Pekka Komsa for helpful conversations.
    This research was supported by an appointment
    (P.S.) to the Intelligence Community Postdoctoral Research Fellowship
    Program at The University of Chicago, administered by Oak Ridge Institute
    for Science and Education through an interagency agreement between the U.S.\ 
    Department of Energy and the Office of the Director of National
    Intelligence.  
    This work (G.G.) was also supported by MICCoM as part of the Computational
    Materials Sciences Program funded by the U.S.\ Department of Energy, Office of
    Science, Basic Energy Sciences, Materials Sciences and Engineering Division through
    Argonne National Laboratory, under contract number DE-AC02-06CH11357.
    An award of computer time was provided by the Innovative and
    Novel Computational Impact on Theory and Experiment (INCITE) program. This
    research used resources of the Argonne Leadership Computing Facility, which
    is a DOE Office of Science User Facility supported under Contract
    DE-AC02-06CH11357 ($GW$ calculations),
    resources of the National Energy
    Research Scientific Computing Center, a DOE Office of Science User Facility
    supported by the Office of Science of the U.S.\ Department of Energy under Contract
    No.\ DE-AC02-05CH11231 (hybrid calculations),
    and resources provided by the
    University of Chicago Research Computing Center.
\end{acknowledgments}

\bibliography{Main_Bibtex}

\begin{thebibliography}{76}%
\makeatletter
\providecommand \@ifxundefined [1]{%
 \@ifx{#1\undefined}
}%
\providecommand \@ifnum [1]{%
 \ifnum #1\expandafter \@firstoftwo
 \else \expandafter \@secondoftwo
 \fi
}%
\providecommand \@ifx [1]{%
 \ifx #1\expandafter \@firstoftwo
 \else \expandafter \@secondoftwo
 \fi
}%
\providecommand \natexlab [1]{#1}%
\providecommand \enquote  [1]{``#1''}%
\providecommand \bibnamefont  [1]{#1}%
\providecommand \bibfnamefont [1]{#1}%
\providecommand \citenamefont [1]{#1}%
\providecommand \href@noop [0]{\@secondoftwo}%
\providecommand \href [0]{\begingroup \@sanitize@url \@href}%
\providecommand \@href[1]{\@@startlink{#1}\@@href}%
\providecommand \@@href[1]{\endgroup#1\@@endlink}%
\providecommand \@sanitize@url [0]{\catcode `\\12\catcode `\$12\catcode
  `\&12\catcode `\#12\catcode `\^12\catcode `\_12\catcode `\%12\relax}%
\providecommand \@@startlink[1]{}%
\providecommand \@@endlink[0]{}%
\providecommand \url  [0]{\begingroup\@sanitize@url \@url }%
\providecommand \@url [1]{\endgroup\@href {#1}{\urlprefix }}%
\providecommand \urlprefix  [0]{URL }%
\providecommand \Eprint [0]{\href }%
\providecommand \doibase [0]{http://dx.doi.org/}%
\providecommand \selectlanguage [0]{\@gobble}%
\providecommand \bibinfo  [0]{\@secondoftwo}%
\providecommand \bibfield  [0]{\@secondoftwo}%
\providecommand \translation [1]{[#1]}%
\providecommand \BibitemOpen [0]{}%
\providecommand \bibitemStop [0]{}%
\providecommand \bibitemNoStop [0]{.\EOS\space}%
\providecommand \EOS [0]{\spacefactor3000\relax}%
\providecommand \BibitemShut  [1]{\csname bibitem#1\endcsname}%
\let\auto@bib@innerbib\@empty
\bibitem [{\citenamefont {Zwanenburg}\ \emph {et~al.}(2013)\citenamefont
  {Zwanenburg}, \citenamefont {Dzurak}, \citenamefont {Morello}, \citenamefont
  {Simmons}, \citenamefont {Hollenberg}, \citenamefont {Klimeck}, \citenamefont
  {Rogge}, \citenamefont {Coppersmith},\ and\ \citenamefont
  {Eriksson}}]{zwanenburg_2013}%
  \BibitemOpen
  \bibfield  {author} {\bibinfo {author} {\bibfnamefont {Floris~A.}\
  \bibnamefont {Zwanenburg}}, \bibinfo {author} {\bibfnamefont {Andrew~S.}\
  \bibnamefont {Dzurak}}, \bibinfo {author} {\bibfnamefont {Andrea}\
  \bibnamefont {Morello}}, \bibinfo {author} {\bibfnamefont {Michelle~Y.}\
  \bibnamefont {Simmons}}, \bibinfo {author} {\bibfnamefont {Lloyd C.~L.}\
  \bibnamefont {Hollenberg}}, \bibinfo {author} {\bibfnamefont {Gerhard}\
  \bibnamefont {Klimeck}}, \bibinfo {author} {\bibfnamefont {Sven}\
  \bibnamefont {Rogge}}, \bibinfo {author} {\bibfnamefont {Susan~N.}\
  \bibnamefont {Coppersmith}}, \ and\ \bibinfo {author} {\bibfnamefont
  {Mark~A.}\ \bibnamefont {Eriksson}},\ }\bibfield  {title} {\enquote {\bibinfo
  {title} {Silicon quantum electronics},}\ }\href {\doibase
  10.1103/RevModPhys.85.961} {\bibfield  {journal} {\bibinfo  {journal} {Rev.
  Mod. Phys.}\ }\textbf {\bibinfo {volume} {85}},\ \bibinfo {pages} {961--1019}
  (\bibinfo {year} {2013})}\BibitemShut {NoStop}%
\bibitem [{\citenamefont {Lyding}\ \emph {et~al.}(1994)\citenamefont {Lyding},
  \citenamefont {Shen}, \citenamefont {Hubacek}, \citenamefont {Tucker},\ and\
  \citenamefont {Abeln}}]{lyding_1994}%
  \BibitemOpen
  \bibfield  {author} {\bibinfo {author} {\bibfnamefont {J.~W.}\ \bibnamefont
  {Lyding}}, \bibinfo {author} {\bibfnamefont {T.-C.}\ \bibnamefont {Shen}},
  \bibinfo {author} {\bibfnamefont {J.~S.}\ \bibnamefont {Hubacek}}, \bibinfo
  {author} {\bibfnamefont {J.~R.}\ \bibnamefont {Tucker}}, \ and\ \bibinfo
  {author} {\bibfnamefont {G.~C.}\ \bibnamefont {Abeln}},\ }\bibfield  {title}
  {\enquote {\bibinfo {title} {Nanoscale patterning and oxidation of
  {H}-passivated {Si(100)-2x1} surfaces with an ultrahigh vacuum scanning
  tunneling microscope},}\ }\href {\doibase 10.1063/1.111722} {\bibfield
  {journal} {\bibinfo  {journal} {Appl. Phys. Lett.}\ }\textbf {\bibinfo
  {volume} {64}},\ \bibinfo {pages} {2010--2012} (\bibinfo {year}
  {1994})}\BibitemShut {NoStop}%
\bibitem [{\citenamefont {Haider}\ \emph {et~al.}(2009)\citenamefont {Haider},
  \citenamefont {Pitters}, \citenamefont {{DiLabio}}, \citenamefont {Livadaru},
  \citenamefont {Mutus},\ and\ \citenamefont {Wolkow}}]{haider_2009}%
  \BibitemOpen
  \bibfield  {author} {\bibinfo {author} {\bibfnamefont {M.~Baseer}\
  \bibnamefont {Haider}}, \bibinfo {author} {\bibfnamefont {Jason~L.}\
  \bibnamefont {Pitters}}, \bibinfo {author} {\bibfnamefont {Gino~A.}\
  \bibnamefont {{DiLabio}}}, \bibinfo {author} {\bibfnamefont {Lucian}\
  \bibnamefont {Livadaru}}, \bibinfo {author} {\bibfnamefont {Josh~Y.}\
  \bibnamefont {Mutus}}, \ and\ \bibinfo {author} {\bibfnamefont {Robert~A.}\
  \bibnamefont {Wolkow}},\ }\bibfield  {title} {\enquote {\bibinfo {title}
  {Controlled coupling and occupation of silicon atomic quantum dots at room
  temperature},}\ }\href {\doibase 10.1103/PhysRevLett.102.046805} {\bibfield
  {journal} {\bibinfo  {journal} {Phys. Rev. Lett.}\ }\textbf {\bibinfo
  {volume} {102}},\ \bibinfo {pages} {046805} (\bibinfo {year}
  {2009})}\BibitemShut {NoStop}%
\bibitem [{\citenamefont {Pitters}\ \emph {et~al.}(2011)\citenamefont
  {Pitters}, \citenamefont {Livadaru}, \citenamefont {Haider},\ and\
  \citenamefont {Wolkow}}]{pitters_2011}%
  \BibitemOpen
  \bibfield  {author} {\bibinfo {author} {\bibfnamefont {Jason~L.}\
  \bibnamefont {Pitters}}, \bibinfo {author} {\bibfnamefont {Lucian}\
  \bibnamefont {Livadaru}}, \bibinfo {author} {\bibfnamefont {M.~Baseer}\
  \bibnamefont {Haider}}, \ and\ \bibinfo {author} {\bibfnamefont {Robert~A.}\
  \bibnamefont {Wolkow}},\ }\bibfield  {title} {\enquote {\bibinfo {title}
  {Tunnel coupled dangling bond structures on hydrogen terminated silicon
  surfaces},}\ }\href {\doibase 10.1063/1.3514896} {\bibfield  {journal}
  {\bibinfo  {journal} {J. Chem. Phys.}\ }\textbf {\bibinfo {volume} {134}},\
  \bibinfo {pages} {064712} (\bibinfo {year} {2011})}\BibitemShut {NoStop}%
\bibitem [{\citenamefont {Bellec}\ \emph {et~al.}(2013)\citenamefont {Bellec},
  \citenamefont {Chaput}, \citenamefont {Dujardin}, \citenamefont {Riedel},
  \citenamefont {Stauffer},\ and\ \citenamefont {Sonnet}}]{bellec_2013}%
  \BibitemOpen
  \bibfield  {author} {\bibinfo {author} {\bibfnamefont {Amandine}\
  \bibnamefont {Bellec}}, \bibinfo {author} {\bibfnamefont {Laurent}\
  \bibnamefont {Chaput}}, \bibinfo {author} {\bibfnamefont {G{\'e}rald}\
  \bibnamefont {Dujardin}}, \bibinfo {author} {\bibfnamefont {Damien}\
  \bibnamefont {Riedel}}, \bibinfo {author} {\bibfnamefont {Louise}\
  \bibnamefont {Stauffer}}, \ and\ \bibinfo {author} {\bibfnamefont {Philippe}\
  \bibnamefont {Sonnet}},\ }\bibfield  {title} {\enquote {\bibinfo {title}
  {Reversible charge storage in a single silicon atom},}\ }\href {\doibase
  10.1103/PhysRevB.88.241406} {\bibfield  {journal} {\bibinfo  {journal} {Phys.
  Rev. B}\ }\textbf {\bibinfo {volume} {88}},\ \bibinfo {pages} {241406}
  (\bibinfo {year} {2013})}\BibitemShut {NoStop}%
\bibitem [{\citenamefont {Schofield}\ \emph {et~al.}(2013)\citenamefont
  {Schofield}, \citenamefont {Studer}, \citenamefont {Hirjibehedin},
  \citenamefont {Curson}, \citenamefont {Aeppli},\ and\ \citenamefont
  {Bowler}}]{schofield_2013}%
  \BibitemOpen
  \bibfield  {author} {\bibinfo {author} {\bibfnamefont {S.~R.}\ \bibnamefont
  {Schofield}}, \bibinfo {author} {\bibfnamefont {P.}~\bibnamefont {Studer}},
  \bibinfo {author} {\bibfnamefont {C.~F.}\ \bibnamefont {Hirjibehedin}},
  \bibinfo {author} {\bibfnamefont {N.~J.}\ \bibnamefont {Curson}}, \bibinfo
  {author} {\bibfnamefont {G.}~\bibnamefont {Aeppli}}, \ and\ \bibinfo {author}
  {\bibfnamefont {D.~R.}\ \bibnamefont {Bowler}},\ }\bibfield  {title}
  {\enquote {\bibinfo {title} {Quantum engineering at the silicon surface using
  dangling bonds},}\ }\href {\doibase 10.1038/ncomms2679} {\bibfield  {journal}
  {\bibinfo  {journal} {Nat. Commun.}\ }\textbf {\bibinfo {volume} {4}},\
  \bibinfo {pages} {1649} (\bibinfo {year} {2013})}\BibitemShut {NoStop}%
\bibitem [{\citenamefont {Ye}\ \emph {et~al.}(2013)\citenamefont {Ye},
  \citenamefont {Min}, \citenamefont {Pe{\~n}a~Martin}, \citenamefont
  {Rockett}, \citenamefont {Aluru},\ and\ \citenamefont {Lyding}}]{ye_2013}%
  \BibitemOpen
  \bibfield  {author} {\bibinfo {author} {\bibfnamefont {Wei}\ \bibnamefont
  {Ye}}, \bibinfo {author} {\bibfnamefont {Kyoungmin}\ \bibnamefont {Min}},
  \bibinfo {author} {\bibfnamefont {Pamela}\ \bibnamefont {Pe{\~n}a~Martin}},
  \bibinfo {author} {\bibfnamefont {Angus~A.}\ \bibnamefont {Rockett}},
  \bibinfo {author} {\bibfnamefont {N.~R.}\ \bibnamefont {Aluru}}, \ and\
  \bibinfo {author} {\bibfnamefont {Joseph~W.}\ \bibnamefont {Lyding}},\
  }\bibfield  {title} {\enquote {\bibinfo {title} {Scanning tunneling
  spectroscopy and density functional calculation of silicon dangling bonds on
  the {Si}(100)-2x1:{H} surface},}\ }\href {\doibase
  10.1016/j.susc.2012.11.015} {\bibfield  {journal} {\bibinfo  {journal} {Surf.
  Sci.}\ }\textbf {\bibinfo {volume} {609}},\ \bibinfo {pages} {147--151}
  (\bibinfo {year} {2013})}\BibitemShut {NoStop}%
\bibitem [{\citenamefont {Taucer}\ \emph {et~al.}(2014)\citenamefont {Taucer},
  \citenamefont {Livadaru}, \citenamefont {Piva}, \citenamefont {Achal},
  \citenamefont {Labidi}, \citenamefont {Pitters},\ and\ \citenamefont
  {Wolkow}}]{taucer_2014}%
  \BibitemOpen
  \bibfield  {author} {\bibinfo {author} {\bibfnamefont {Marco}\ \bibnamefont
  {Taucer}}, \bibinfo {author} {\bibfnamefont {Lucian}\ \bibnamefont
  {Livadaru}}, \bibinfo {author} {\bibfnamefont {Paul~G.}\ \bibnamefont
  {Piva}}, \bibinfo {author} {\bibfnamefont {Roshan}\ \bibnamefont {Achal}},
  \bibinfo {author} {\bibfnamefont {Hatem}\ \bibnamefont {Labidi}}, \bibinfo
  {author} {\bibfnamefont {Jason~L.}\ \bibnamefont {Pitters}}, \ and\ \bibinfo
  {author} {\bibfnamefont {Robert~A.}\ \bibnamefont {Wolkow}},\ }\bibfield
  {title} {\enquote {\bibinfo {title} {Single-electron dynamics of an atomic
  silicon quantum dot on the {H-Si}(100)-$(2{\times}1)$ surface},}\ }\href
  {\doibase 10.1103/PhysRevLett.112.256801} {\bibfield  {journal} {\bibinfo
  {journal} {Phys. Rev. Lett.}\ }\textbf {\bibinfo {volume} {112}},\ \bibinfo
  {pages} {256801} (\bibinfo {year} {2014})}\BibitemShut {NoStop}%
\bibitem [{\citenamefont {Kolmer}\ \emph {et~al.}(2014)\citenamefont {Kolmer},
  \citenamefont {Godlewski}, \citenamefont {Zuzak}, \citenamefont {Wojtaszek},
  \citenamefont {Rauer}, \citenamefont {Thuaire}, \citenamefont {Hartmann},
  \citenamefont {Moriceau}, \citenamefont {Joachim},\ and\ \citenamefont
  {Szymonski}}]{kolmer_2014}%
  \BibitemOpen
  \bibfield  {author} {\bibinfo {author} {\bibfnamefont {Marek}\ \bibnamefont
  {Kolmer}}, \bibinfo {author} {\bibfnamefont {Szymon}\ \bibnamefont
  {Godlewski}}, \bibinfo {author} {\bibfnamefont {Rafal}\ \bibnamefont
  {Zuzak}}, \bibinfo {author} {\bibfnamefont {Mateusz}\ \bibnamefont
  {Wojtaszek}}, \bibinfo {author} {\bibfnamefont {Caroline}\ \bibnamefont
  {Rauer}}, \bibinfo {author} {\bibfnamefont {Aur{\'e}lie}\ \bibnamefont
  {Thuaire}}, \bibinfo {author} {\bibfnamefont {Jean-Michel}\ \bibnamefont
  {Hartmann}}, \bibinfo {author} {\bibfnamefont {Hubert}\ \bibnamefont
  {Moriceau}}, \bibinfo {author} {\bibfnamefont {Christian}\ \bibnamefont
  {Joachim}}, \ and\ \bibinfo {author} {\bibfnamefont {Marek}\ \bibnamefont
  {Szymonski}},\ }\bibfield  {title} {\enquote {\bibinfo {title} {Atomic scale
  fabrication of dangling bond structures on hydrogen passivated {Si}(001)
  wafers processed and nanopackaged in a clean room environment},}\ }\href
  {\doibase 10.1016/j.apsusc.2013.09.124} {\bibfield  {journal} {\bibinfo
  {journal} {Appl. Surf. Sci.}\ }\textbf {\bibinfo {volume} {288}},\ \bibinfo
  {pages} {83--89} (\bibinfo {year} {2014})}\BibitemShut {NoStop}%
\bibitem [{\citenamefont {Labidi}\ \emph {et~al.}(2015)\citenamefont {Labidi},
  \citenamefont {Taucer}, \citenamefont {Rashidi}, \citenamefont {Koleini},
  \citenamefont {Livadaru}, \citenamefont {Pitters}, \citenamefont {{Martin
  Cloutier}}, \citenamefont {Salomons},\ and\ \citenamefont
  {Wolkow}}]{labidi_2015}%
  \BibitemOpen
  \bibfield  {author} {\bibinfo {author} {\bibfnamefont {Hatem}\ \bibnamefont
  {Labidi}}, \bibinfo {author} {\bibfnamefont {Marco}\ \bibnamefont {Taucer}},
  \bibinfo {author} {\bibfnamefont {Mohammad}\ \bibnamefont {Rashidi}},
  \bibinfo {author} {\bibfnamefont {Mohammad}\ \bibnamefont {Koleini}},
  \bibinfo {author} {\bibfnamefont {Lucian}\ \bibnamefont {Livadaru}}, \bibinfo
  {author} {\bibfnamefont {Jason}\ \bibnamefont {Pitters}}, \bibinfo {author}
  {\bibnamefont {{Martin Cloutier}}}, \bibinfo {author} {\bibfnamefont {Mark}\
  \bibnamefont {Salomons}}, \ and\ \bibinfo {author} {\bibfnamefont
  {Robert~A.}\ \bibnamefont {Wolkow}},\ }\bibfield  {title} {\enquote {\bibinfo
  {title} {Scanning tunneling spectroscopy reveals a silicon dangling bond
  charge state transition},}\ }\href {\doibase 10.1088/1367-2630/17/7/073023}
  {\bibfield  {journal} {\bibinfo  {journal} {New J. Phys.}\ }\textbf {\bibinfo
  {volume} {17}},\ \bibinfo {pages} {073023} (\bibinfo {year}
  {2015})}\BibitemShut {NoStop}%
\bibitem [{\citenamefont {Kawai}\ \emph {et~al.}(2016)\citenamefont {Kawai},
  \citenamefont {Neucheva}, \citenamefont {Yap}, \citenamefont {Joachim},\ and\
  \citenamefont {Saeys}}]{kawai_2016}%
  \BibitemOpen
  \bibfield  {author} {\bibinfo {author} {\bibfnamefont {Hiroyo}\ \bibnamefont
  {Kawai}}, \bibinfo {author} {\bibfnamefont {Olga}\ \bibnamefont {Neucheva}},
  \bibinfo {author} {\bibfnamefont {Tiong~Leh}\ \bibnamefont {Yap}}, \bibinfo
  {author} {\bibfnamefont {Christian}\ \bibnamefont {Joachim}}, \ and\ \bibinfo
  {author} {\bibfnamefont {Mark}\ \bibnamefont {Saeys}},\ }\bibfield  {title}
  {\enquote {\bibinfo {title} {Electronic characterization of a single dangling
  bond on n- and p-type {Si(001)-(2x1):H}},}\ }\href {\doibase
  10.1016/j.susc.2015.11.001} {\bibfield  {journal} {\bibinfo  {journal} {Surf.
  Sci.}\ }\textbf {\bibinfo {volume} {645}},\ \bibinfo {pages} {88--92}
  (\bibinfo {year} {2016})}\BibitemShut {NoStop}%
\bibitem [{\citenamefont {Rashidi}\ \emph {et~al.}(2016)\citenamefont
  {Rashidi}, \citenamefont {Taucer}, \citenamefont {Ozfidan}, \citenamefont
  {Lloyd}, \citenamefont {Koleini}, \citenamefont {Labidi}, \citenamefont
  {Pitters}, \citenamefont {Maciejko},\ and\ \citenamefont
  {Wolkow}}]{rashidi_2016}%
  \BibitemOpen
  \bibfield  {author} {\bibinfo {author} {\bibfnamefont {Mohammad}\
  \bibnamefont {Rashidi}}, \bibinfo {author} {\bibfnamefont {Marco}\
  \bibnamefont {Taucer}}, \bibinfo {author} {\bibfnamefont {Isil}\ \bibnamefont
  {Ozfidan}}, \bibinfo {author} {\bibfnamefont {Erika}\ \bibnamefont {Lloyd}},
  \bibinfo {author} {\bibfnamefont {Mohammad}\ \bibnamefont {Koleini}},
  \bibinfo {author} {\bibfnamefont {Hatem}\ \bibnamefont {Labidi}}, \bibinfo
  {author} {\bibfnamefont {Jason~L.}\ \bibnamefont {Pitters}}, \bibinfo
  {author} {\bibfnamefont {Joseph}\ \bibnamefont {Maciejko}}, \ and\ \bibinfo
  {author} {\bibfnamefont {Robert~A.}\ \bibnamefont {Wolkow}},\ }\bibfield
  {title} {\enquote {\bibinfo {title} {Time-resolved imaging of negative
  differential resistance on the atomic scale},}\ }\href {\doibase
  10.1103/PhysRevLett.117.276805} {\bibfield  {journal} {\bibinfo  {journal}
  {Phys. Rev. Lett.}\ }\textbf {\bibinfo {volume} {117}},\ \bibinfo {pages}
  {276805} (\bibinfo {year} {2016})}\BibitemShut {NoStop}%
\bibitem [{\citenamefont {Hitosugi}\ \emph {et~al.}(1998)\citenamefont
  {Hitosugi}, \citenamefont {Hashizume}, \citenamefont {Heike}, \citenamefont
  {Wada}, \citenamefont {Watanabe}, \citenamefont {Hasegawa},\ and\
  \citenamefont {Kitazawa}}]{hitosugi_1998}%
  \BibitemOpen
  \bibfield  {author} {\bibinfo {author} {\bibfnamefont {Taro}\ \bibnamefont
  {Hitosugi}}, \bibinfo {author} {\bibfnamefont {T.}~\bibnamefont {Hashizume}},
  \bibinfo {author} {\bibfnamefont {S.}~\bibnamefont {Heike}}, \bibinfo
  {author} {\bibfnamefont {Y.}~\bibnamefont {Wada}}, \bibinfo {author}
  {\bibfnamefont {S.}~\bibnamefont {Watanabe}}, \bibinfo {author}
  {\bibfnamefont {T.}~\bibnamefont {Hasegawa}}, \ and\ \bibinfo {author}
  {\bibfnamefont {K.}~\bibnamefont {Kitazawa}},\ }\bibfield  {title} {\enquote
  {\bibinfo {title} {Scanning tunnelling spectroscopy of dangling-bond wires
  fabricated on the {Si}(100)-2${\times}$1-{H} surface},}\ }\href {\doibase
  10.1007/s003390051224} {\bibfield  {journal} {\bibinfo  {journal} {Appl.
  Phys. A}\ }\textbf {\bibinfo {volume} {66}},\ \bibinfo {pages} {S695--S699}
  (\bibinfo {year} {1998})}\BibitemShut {NoStop}%
\bibitem [{\citenamefont {Soukiassian}\ \emph {et~al.}(2003)\citenamefont
  {Soukiassian}, \citenamefont {Mayne}, \citenamefont {Carbone},\ and\
  \citenamefont {Dujardin}}]{soukiassian_2003}%
  \BibitemOpen
  \bibfield  {author} {\bibinfo {author} {\bibfnamefont {Laetitia}\
  \bibnamefont {Soukiassian}}, \bibinfo {author} {\bibfnamefont {Andrew~J.}\
  \bibnamefont {Mayne}}, \bibinfo {author} {\bibfnamefont {Marilena}\
  \bibnamefont {Carbone}}, \ and\ \bibinfo {author} {\bibfnamefont
  {G{\'{e}}rald}\ \bibnamefont {Dujardin}},\ }\bibfield  {title} {\enquote
  {\bibinfo {title} {Atomic wire fabrication by {STM} induced hydrogen
  desorption},}\ }\href {\doibase 10.1016/S0039-6028(02)02620-1} {\bibfield
  {journal} {\bibinfo  {journal} {Surf. Sci.}\ }\bibinfo {series} {Proceedings
  of the Ninth International Workshop on Desorption Induced by Electronic
  Transitions},\ \textbf {\bibinfo {volume} {528}},\ \bibinfo {pages}
  {121--126} (\bibinfo {year} {2003})}\BibitemShut {NoStop}%
\bibitem [{\citenamefont {Mayne}\ \emph {et~al.}(2006)\citenamefont {Mayne},
  \citenamefont {Riedel}, \citenamefont {Comtet},\ and\ \citenamefont
  {Dujardin}}]{mayne_2006}%
  \BibitemOpen
  \bibfield  {author} {\bibinfo {author} {\bibfnamefont {A.~J.}\ \bibnamefont
  {Mayne}}, \bibinfo {author} {\bibfnamefont {D.}~\bibnamefont {Riedel}},
  \bibinfo {author} {\bibfnamefont {G.}~\bibnamefont {Comtet}}, \ and\ \bibinfo
  {author} {\bibfnamefont {G.}~\bibnamefont {Dujardin}},\ }\bibfield  {title}
  {\enquote {\bibinfo {title} {Atomic-scale studies of hydrogenated
  semiconductor surfaces},}\ }\href {\doibase 10.1016/j.progsurf.2006.01.001}
  {\bibfield  {journal} {\bibinfo  {journal} {Prog. Surf. Sci.}\ }\textbf
  {\bibinfo {volume} {81}},\ \bibinfo {pages} {1--51} (\bibinfo {year}
  {2006})}\BibitemShut {NoStop}%
\bibitem [{\citenamefont {Livadaru}\ \emph {et~al.}(2010)\citenamefont
  {Livadaru}, \citenamefont {Xue}, \citenamefont {Shaterzadeh-Yazdi},
  \citenamefont {{DiLabio}}, \citenamefont {Mutus}, \citenamefont {Pitters},
  \citenamefont {Sanders},\ and\ \citenamefont {Wolkow}}]{livadaru_2010}%
  \BibitemOpen
  \bibfield  {author} {\bibinfo {author} {\bibfnamefont {Lucian}\ \bibnamefont
  {Livadaru}}, \bibinfo {author} {\bibfnamefont {Peng}\ \bibnamefont {Xue}},
  \bibinfo {author} {\bibfnamefont {Zahra}\ \bibnamefont {Shaterzadeh-Yazdi}},
  \bibinfo {author} {\bibfnamefont {Gino~A}\ \bibnamefont {{DiLabio}}},
  \bibinfo {author} {\bibfnamefont {Josh}\ \bibnamefont {Mutus}}, \bibinfo
  {author} {\bibfnamefont {Jason~L}\ \bibnamefont {Pitters}}, \bibinfo {author}
  {\bibfnamefont {Barry~C}\ \bibnamefont {Sanders}}, \ and\ \bibinfo {author}
  {\bibfnamefont {Robert~A}\ \bibnamefont {Wolkow}},\ }\bibfield  {title}
  {\enquote {\bibinfo {title} {Dangling-bond charge qubit on a silicon
  surface},}\ }\href {\doibase 10.1088/1367-2630/12/8/083018} {\bibfield
  {journal} {\bibinfo  {journal} {New J. Phys.}\ }\textbf {\bibinfo {volume}
  {12}},\ \bibinfo {pages} {083018} (\bibinfo {year} {2010})}\BibitemShut
  {NoStop}%
\bibitem [{\citenamefont {Ample}\ \emph {et~al.}(2011)\citenamefont {Ample},
  \citenamefont {Duchemin}, \citenamefont {Hliwa},\ and\ \citenamefont
  {Joachim}}]{ample_2011}%
  \BibitemOpen
  \bibfield  {author} {\bibinfo {author} {\bibfnamefont {F.}~\bibnamefont
  {Ample}}, \bibinfo {author} {\bibfnamefont {I.}~\bibnamefont {Duchemin}},
  \bibinfo {author} {\bibfnamefont {M.}~\bibnamefont {Hliwa}}, \ and\ \bibinfo
  {author} {\bibfnamefont {C.}~\bibnamefont {Joachim}},\ }\bibfield  {title}
  {\enquote {\bibinfo {title} {Single {OR} molecule and {OR} atomic circuit
  logic gates interconnected on a {Si(100)H} surface},}\ }\href {\doibase
  10.1088/0953-8984/23/12/125303} {\bibfield  {journal} {\bibinfo  {journal}
  {J. Phys.: Condens. Matter}\ }\textbf {\bibinfo {volume} {23}},\ \bibinfo
  {pages} {125303} (\bibinfo {year} {2011})}\BibitemShut {NoStop}%
\bibitem [{\citenamefont {Kepenekian}\ \emph {et~al.}(2014)\citenamefont
  {Kepenekian}, \citenamefont {Robles}, \citenamefont {Rurali},\ and\
  \citenamefont {Lorente}}]{kepenekian_2014}%
  \BibitemOpen
  \bibfield  {author} {\bibinfo {author} {\bibfnamefont {Mika{\"{e}}l}\
  \bibnamefont {Kepenekian}}, \bibinfo {author} {\bibfnamefont {Roberto}\
  \bibnamefont {Robles}}, \bibinfo {author} {\bibfnamefont {Riccardo}\
  \bibnamefont {Rurali}}, \ and\ \bibinfo {author} {\bibfnamefont
  {Nicol{\'{a}}s}\ \bibnamefont {Lorente}},\ }\bibfield  {title} {\enquote
  {\bibinfo {title} {Spin transport in dangling-bond wires on doped
  {H}-passivated {Si(100)}},}\ }\href {\doibase 10.1088/0957-4484/25/46/465703}
  {\bibfield  {journal} {\bibinfo  {journal} {Nanotechnology}\ }\textbf
  {\bibinfo {volume} {25}},\ \bibinfo {pages} {465703} (\bibinfo {year}
  {2014})}\BibitemShut {NoStop}%
\bibitem [{\citenamefont {Hitosugi}\ \emph {et~al.}(1999)\citenamefont
  {Hitosugi}, \citenamefont {Heike}, \citenamefont {Onogi}, \citenamefont
  {Hashizume}, \citenamefont {Watanabe}, \citenamefont {Li}, \citenamefont
  {Ohno}, \citenamefont {Kawazoe}, \citenamefont {Hasegawa},\ and\
  \citenamefont {Kitazawa}}]{hitosugi_1999}%
  \BibitemOpen
  \bibfield  {author} {\bibinfo {author} {\bibfnamefont {Taro}\ \bibnamefont
  {Hitosugi}}, \bibinfo {author} {\bibfnamefont {S.}~\bibnamefont {Heike}},
  \bibinfo {author} {\bibfnamefont {T.}~\bibnamefont {Onogi}}, \bibinfo
  {author} {\bibfnamefont {T.}~\bibnamefont {Hashizume}}, \bibinfo {author}
  {\bibfnamefont {S.}~\bibnamefont {Watanabe}}, \bibinfo {author}
  {\bibfnamefont {Z.-Q.}\ \bibnamefont {Li}}, \bibinfo {author} {\bibfnamefont
  {K.}~\bibnamefont {Ohno}}, \bibinfo {author} {\bibfnamefont {Y.}~\bibnamefont
  {Kawazoe}}, \bibinfo {author} {\bibfnamefont {T.}~\bibnamefont {Hasegawa}}, \
  and\ \bibinfo {author} {\bibfnamefont {K.}~\bibnamefont {Kitazawa}},\
  }\bibfield  {title} {\enquote {\bibinfo {title} {Jahn-teller distortion in
  dangling-bond linear chains fabricated on a hydrogen-terminated
  {Si}(100)-$2\times1$ surface},}\ }\href {\doibase
  10.1103/PhysRevLett.82.4034} {\bibfield  {journal} {\bibinfo  {journal}
  {Phys. Rev. Lett.}\ }\textbf {\bibinfo {volume} {82}},\ \bibinfo {pages}
  {4034--4037} (\bibinfo {year} {1999})}\BibitemShut {NoStop}%
\bibitem [{\citenamefont {Naydenov}\ and\ \citenamefont
  {Boland}(2013)}]{naydenov_2013}%
  \BibitemOpen
  \bibfield  {author} {\bibinfo {author} {\bibfnamefont {Borislav}\
  \bibnamefont {Naydenov}}\ and\ \bibinfo {author} {\bibfnamefont {John~J.}\
  \bibnamefont {Boland}},\ }\bibfield  {title} {\enquote {\bibinfo {title}
  {Engineering the electronic structure of surface dangling bond nanowires of
  different size and dimensionality},}\ }\href {\doibase
  10.1088/0957-4484/24/27/275202} {\bibfield  {journal} {\bibinfo  {journal}
  {Nanotechnology}\ }\textbf {\bibinfo {volume} {24}},\ \bibinfo {pages}
  {275202} (\bibinfo {year} {2013})}\BibitemShut {NoStop}%
\bibitem [{\citenamefont {Schofield}\ \emph {et~al.}(2003)\citenamefont
  {Schofield}, \citenamefont {Curson}, \citenamefont {Simmons}, \citenamefont
  {Rue{\ss}}, \citenamefont {Hallam}, \citenamefont {Oberbeck},\ and\
  \citenamefont {Clark}}]{schofield_2003}%
  \BibitemOpen
  \bibfield  {author} {\bibinfo {author} {\bibfnamefont {S.~R.}\ \bibnamefont
  {Schofield}}, \bibinfo {author} {\bibfnamefont {N.~J.}\ \bibnamefont
  {Curson}}, \bibinfo {author} {\bibfnamefont {M.~Y.}\ \bibnamefont {Simmons}},
  \bibinfo {author} {\bibfnamefont {F.~J.}\ \bibnamefont {Rue{\ss}}}, \bibinfo
  {author} {\bibfnamefont {T.}~\bibnamefont {Hallam}}, \bibinfo {author}
  {\bibfnamefont {L.}~\bibnamefont {Oberbeck}}, \ and\ \bibinfo {author}
  {\bibfnamefont {R.~G.}\ \bibnamefont {Clark}},\ }\bibfield  {title} {\enquote
  {\bibinfo {title} {Atomically precise placement of single dopants in {Si}},}\
  }\href {\doibase 10.1103/PhysRevLett.91.136104} {\bibfield  {journal}
  {\bibinfo  {journal} {Phys. Rev. Lett.}\ }\textbf {\bibinfo {volume} {91}},\
  \bibinfo {pages} {136104} (\bibinfo {year} {2003})}\BibitemShut {NoStop}%
\bibitem [{\citenamefont {Ruess}\ \emph {et~al.}(2004)\citenamefont {Ruess},
  \citenamefont {Oberbeck}, \citenamefont {Simmons}, \citenamefont {Goh},
  \citenamefont {Hamilton}, \citenamefont {Hallam}, \citenamefont {Schofield},
  \citenamefont {Curson},\ and\ \citenamefont {Clark}}]{ruess_2004}%
  \BibitemOpen
  \bibfield  {author} {\bibinfo {author} {\bibfnamefont {Frank~J.}\
  \bibnamefont {Ruess}}, \bibinfo {author} {\bibfnamefont {Lars}\ \bibnamefont
  {Oberbeck}}, \bibinfo {author} {\bibfnamefont {Michelle~Y.}\ \bibnamefont
  {Simmons}}, \bibinfo {author} {\bibfnamefont {Kuan Eng~J.}\ \bibnamefont
  {Goh}}, \bibinfo {author} {\bibfnamefont {Alex~R.}\ \bibnamefont {Hamilton}},
  \bibinfo {author} {\bibfnamefont {Toby}\ \bibnamefont {Hallam}}, \bibinfo
  {author} {\bibfnamefont {Steven~R.}\ \bibnamefont {Schofield}}, \bibinfo
  {author} {\bibfnamefont {Neil~J.}\ \bibnamefont {Curson}}, \ and\ \bibinfo
  {author} {\bibfnamefont {Robert~G.}\ \bibnamefont {Clark}},\ }\bibfield
  {title} {\enquote {\bibinfo {title} {Toward atomic-scale device fabrication
  in silicon using scanning probe microscopy},}\ }\href {\doibase
  10.1021/nl048808v} {\bibfield  {journal} {\bibinfo  {journal} {Nano Lett.}\
  }\textbf {\bibinfo {volume} {4}},\ \bibinfo {pages} {1969--1973} (\bibinfo
  {year} {2004})}\BibitemShut {NoStop}%
\bibitem [{\citenamefont {Fuechsle}\ \emph {et~al.}(2012)\citenamefont
  {Fuechsle}, \citenamefont {Miwa}, \citenamefont {Mahapatra}, \citenamefont
  {Ryu}, \citenamefont {Lee}, \citenamefont {Warschkow}, \citenamefont
  {Hollenberg}, \citenamefont {Klimeck},\ and\ \citenamefont
  {Simmons}}]{fuechsle_2012}%
  \BibitemOpen
  \bibfield  {author} {\bibinfo {author} {\bibfnamefont {Martin}\ \bibnamefont
  {Fuechsle}}, \bibinfo {author} {\bibfnamefont {Jill~A.}\ \bibnamefont
  {Miwa}}, \bibinfo {author} {\bibfnamefont {Suddhasatta}\ \bibnamefont
  {Mahapatra}}, \bibinfo {author} {\bibfnamefont {Hoon}\ \bibnamefont {Ryu}},
  \bibinfo {author} {\bibfnamefont {Sunhee}\ \bibnamefont {Lee}}, \bibinfo
  {author} {\bibfnamefont {Oliver}\ \bibnamefont {Warschkow}}, \bibinfo
  {author} {\bibfnamefont {Lloyd C.~L.}\ \bibnamefont {Hollenberg}}, \bibinfo
  {author} {\bibfnamefont {Gerhard}\ \bibnamefont {Klimeck}}, \ and\ \bibinfo
  {author} {\bibfnamefont {Michelle~Y.}\ \bibnamefont {Simmons}},\ }\bibfield
  {title} {\enquote {\bibinfo {title} {A single-atom transistor},}\ }\href
  {\doibase 10.1038/nnano.2012.21} {\bibfield  {journal} {\bibinfo  {journal}
  {Nat. Nano.}\ }\textbf {\bibinfo {volume} {7}},\ \bibinfo {pages} {242--246}
  (\bibinfo {year} {2012})}\BibitemShut {NoStop}%
\bibitem [{\citenamefont {Weber}\ \emph {et~al.}(2012)\citenamefont {Weber},
  \citenamefont {Mahapatra}, \citenamefont {Ryu}, \citenamefont {Lee},
  \citenamefont {Fuhrer}, \citenamefont {Reusch}, \citenamefont {Thompson},
  \citenamefont {Lee}, \citenamefont {Klimeck}, \citenamefont {Hollenberg},\
  and\ \citenamefont {Simmons}}]{weber_2012}%
  \BibitemOpen
  \bibfield  {author} {\bibinfo {author} {\bibfnamefont {B.}~\bibnamefont
  {Weber}}, \bibinfo {author} {\bibfnamefont {S.}~\bibnamefont {Mahapatra}},
  \bibinfo {author} {\bibfnamefont {H.}~\bibnamefont {Ryu}}, \bibinfo {author}
  {\bibfnamefont {S.}~\bibnamefont {Lee}}, \bibinfo {author} {\bibfnamefont
  {A.}~\bibnamefont {Fuhrer}}, \bibinfo {author} {\bibfnamefont {T.~C.~G.}\
  \bibnamefont {Reusch}}, \bibinfo {author} {\bibfnamefont {D.~L.}\
  \bibnamefont {Thompson}}, \bibinfo {author} {\bibfnamefont {W.~C.~T.}\
  \bibnamefont {Lee}}, \bibinfo {author} {\bibfnamefont {G.}~\bibnamefont
  {Klimeck}}, \bibinfo {author} {\bibfnamefont {L.~C.~L.}\ \bibnamefont
  {Hollenberg}}, \ and\ \bibinfo {author} {\bibfnamefont {M.~Y.}\ \bibnamefont
  {Simmons}},\ }\bibfield  {title} {\enquote {\bibinfo {title} {Ohm's law
  survives to the atomic scale},}\ }\href {\doibase 10.1126/science.1214319}
  {\bibfield  {journal} {\bibinfo  {journal} {Science}\ }\textbf {\bibinfo
  {volume} {335}},\ \bibinfo {pages} {64--67} (\bibinfo {year}
  {2012})}\BibitemShut {NoStop}%
\bibitem [{\citenamefont {Morley}(2014)}]{morley_2014}%
  \BibitemOpen
  \bibfield  {author} {\bibinfo {author} {\bibfnamefont {Gavin~W.}\
  \bibnamefont {Morley}},\ }\bibfield  {title} {\enquote {\bibinfo {title}
  {Chapter 3: Towards spintronic quantum technologies with dopants in
  silicon},}\ }in\ \href@noop {} {\emph {\bibinfo {booktitle} {Electron
  Paramagnetic Resonance}}},\ Vol.~\bibinfo {volume} {24}\ (\bibinfo
  {publisher} {Royal Society of Chemistry},\ \bibinfo {year} {2014})\ pp.\
  \bibinfo {pages} {62--76}\BibitemShut {NoStop}%
\bibitem [{\citenamefont {Hohenberg}\ and\ \citenamefont
  {Kohn}(1964)}]{hohenberg_1964}%
  \BibitemOpen
  \bibfield  {author} {\bibinfo {author} {\bibfnamefont {P.}~\bibnamefont
  {Hohenberg}}\ and\ \bibinfo {author} {\bibfnamefont {W.}~\bibnamefont
  {Kohn}},\ }\bibfield  {title} {\enquote {\bibinfo {title} {Inhomogeneous
  electron gas},}\ }\href {\doibase 10.1103/PhysRev.136.B864} {\bibfield
  {journal} {\bibinfo  {journal} {Phys. Rev.}\ }\textbf {\bibinfo {volume}
  {136}},\ \bibinfo {pages} {B864--B871} (\bibinfo {year} {1964})}\BibitemShut
  {NoStop}%
\bibitem [{\citenamefont {Kohn}\ and\ \citenamefont {Sham}(1965)}]{kohn_1965}%
  \BibitemOpen
  \bibfield  {author} {\bibinfo {author} {\bibfnamefont {W.}~\bibnamefont
  {Kohn}}\ and\ \bibinfo {author} {\bibfnamefont {L.~J.}\ \bibnamefont
  {Sham}},\ }\bibfield  {title} {\enquote {\bibinfo {title} {Self-consistent
  equations including exchange and correlation effects},}\ }\href {\doibase
  10.1103/PhysRev.140.A1133} {\bibfield  {journal} {\bibinfo  {journal} {Phys.
  Rev.}\ }\textbf {\bibinfo {volume} {140}},\ \bibinfo {pages} {A1133--A1138}
  (\bibinfo {year} {1965})}\BibitemShut {NoStop}%
\bibitem [{\citenamefont {Troullier}\ and\ \citenamefont
  {Martins}(1991)}]{troullier_1991}%
  \BibitemOpen
  \bibfield  {author} {\bibinfo {author} {\bibfnamefont {N.}~\bibnamefont
  {Troullier}}\ and\ \bibinfo {author} {\bibfnamefont {Jos{\'e}~Lu{\'i}s}\
  \bibnamefont {Martins}},\ }\bibfield  {title} {\enquote {\bibinfo {title}
  {Efficient pseudopotentials for plane-wave calculations},}\ }\href {\doibase
  10.1103/PhysRevB.43.1993} {\bibfield  {journal} {\bibinfo  {journal} {Phys.
  Rev. B}\ }\textbf {\bibinfo {volume} {43}},\ \bibinfo {pages} {1993--2006}
  (\bibinfo {year} {1991})}\BibitemShut {NoStop}%
\bibitem [{\citenamefont {Ceresoli}()}]{ceresoli_2016}%
  \BibitemOpen
  \bibfield  {author} {\bibinfo {author} {\bibfnamefont {Davide}\ \bibnamefont
  {Ceresoli}},\ }\href
  {https://sites.google.com/site/dceresoli/pseudopotentials} {\enquote
  {\bibinfo {title} {Pseudopotentials},}\ }\bibinfo {note}
  {{https://sites.google.com/site/dceresoli/pseudopotentials}, (accessed May
  20, 2016)}\BibitemShut {NoStop}%
\bibitem [{\citenamefont {Giannozzi}\ \emph {et~al.}(2009)\citenamefont
  {Giannozzi}, \citenamefont {Baroni}, \citenamefont {Bonini}, \citenamefont
  {Calandra}, \citenamefont {Car}, \citenamefont {Cavazzoni}, \citenamefont
  {Ceresoli}, \citenamefont {Chiarotti}, \citenamefont {Cococcioni},
  \citenamefont {Dabo}, \citenamefont {Dal~Corso}, \citenamefont {Gironcoli},
  \citenamefont {Fabris}, \citenamefont {Fratesi}, \citenamefont {Gebauer},
  \citenamefont {Gerstmann}, \citenamefont {Gougoussis}, \citenamefont
  {Kokalj}, \citenamefont {Lazzeri}, \citenamefont {Martin-Samos},
  \citenamefont {Marzari}, \citenamefont {Mauri}, \citenamefont {Mazzarello},
  \citenamefont {Paolini}, \citenamefont {Pasquarello}, \citenamefont
  {Paulatto}, \citenamefont {Sbraccia}, \citenamefont {Scandolo}, \citenamefont
  {Sclauzero}, \citenamefont {Seitsonen}, \citenamefont {Smogunov},
  \citenamefont {Umari},\ and\ \citenamefont {Wentzcovitch}}]{giannozzi_2009}%
  \BibitemOpen
  \bibfield  {author} {\bibinfo {author} {\bibfnamefont {Paolo}\ \bibnamefont
  {Giannozzi}}, \bibinfo {author} {\bibfnamefont {Stefano}\ \bibnamefont
  {Baroni}}, \bibinfo {author} {\bibfnamefont {Nicola}\ \bibnamefont {Bonini}},
  \bibinfo {author} {\bibfnamefont {Matteo}\ \bibnamefont {Calandra}}, \bibinfo
  {author} {\bibfnamefont {Roberto}\ \bibnamefont {Car}}, \bibinfo {author}
  {\bibfnamefont {Carlo}\ \bibnamefont {Cavazzoni}}, \bibinfo {author}
  {\bibfnamefont {Davide}\ \bibnamefont {Ceresoli}}, \bibinfo {author}
  {\bibfnamefont {Guido~L.}\ \bibnamefont {Chiarotti}}, \bibinfo {author}
  {\bibfnamefont {Matteo}\ \bibnamefont {Cococcioni}}, \bibinfo {author}
  {\bibfnamefont {Ismaila}\ \bibnamefont {Dabo}}, \bibinfo {author}
  {\bibfnamefont {Andrea}\ \bibnamefont {Dal~Corso}}, \bibinfo {author}
  {\bibfnamefont {Stefano~de}\ \bibnamefont {Gironcoli}}, \bibinfo {author}
  {\bibfnamefont {Stefano}\ \bibnamefont {Fabris}}, \bibinfo {author}
  {\bibfnamefont {Guido}\ \bibnamefont {Fratesi}}, \bibinfo {author}
  {\bibfnamefont {Ralph}\ \bibnamefont {Gebauer}}, \bibinfo {author}
  {\bibfnamefont {Uwe}\ \bibnamefont {Gerstmann}}, \bibinfo {author}
  {\bibfnamefont {Christos}\ \bibnamefont {Gougoussis}}, \bibinfo {author}
  {\bibfnamefont {Anton}\ \bibnamefont {Kokalj}}, \bibinfo {author}
  {\bibfnamefont {Michele}\ \bibnamefont {Lazzeri}}, \bibinfo {author}
  {\bibfnamefont {Layla}\ \bibnamefont {Martin-Samos}}, \bibinfo {author}
  {\bibfnamefont {Nicola}\ \bibnamefont {Marzari}}, \bibinfo {author}
  {\bibfnamefont {Francesco}\ \bibnamefont {Mauri}}, \bibinfo {author}
  {\bibfnamefont {Riccardo}\ \bibnamefont {Mazzarello}}, \bibinfo {author}
  {\bibfnamefont {Stefano}\ \bibnamefont {Paolini}}, \bibinfo {author}
  {\bibfnamefont {Alfredo}\ \bibnamefont {Pasquarello}}, \bibinfo {author}
  {\bibfnamefont {Lorenzo}\ \bibnamefont {Paulatto}}, \bibinfo {author}
  {\bibfnamefont {Carlo}\ \bibnamefont {Sbraccia}}, \bibinfo {author}
  {\bibfnamefont {Sandro}\ \bibnamefont {Scandolo}}, \bibinfo {author}
  {\bibfnamefont {Gabriele}\ \bibnamefont {Sclauzero}}, \bibinfo {author}
  {\bibfnamefont {Ari~P.}\ \bibnamefont {Seitsonen}}, \bibinfo {author}
  {\bibfnamefont {Alexander}\ \bibnamefont {Smogunov}}, \bibinfo {author}
  {\bibfnamefont {Paolo}\ \bibnamefont {Umari}}, \ and\ \bibinfo {author}
  {\bibfnamefont {Renata~M.}\ \bibnamefont {Wentzcovitch}},\ }\bibfield
  {title} {\enquote {\bibinfo {title} {{QUANTUM} {ESPRESSO}: a modular and
  open-source software project for quantum simulations of materials},}\ }\href
  {\doibase 10.1088/0953-8984/21/39/395502} {\bibfield  {journal} {\bibinfo
  {journal} {J. Phys.: Condens. Matter}\ }\textbf {\bibinfo {volume} {21}},\
  \bibinfo {pages} {395502} (\bibinfo {year} {2009})}\BibitemShut {NoStop}%
\bibitem [{qe_()}]{qe_web}%
  \BibitemOpen
  \href {http://www.quantum-espresso.org/} {\enquote {\bibinfo {title}
  {{QUANTUM ESPRESSO}},}\ }\bibinfo {note} {{http://www.quantum-espresso.org},
  (accessed Jan 26, 2016)}\BibitemShut {NoStop}%
\bibitem [{\citenamefont {Perdew}\ \emph {et~al.}(1996)\citenamefont {Perdew},
  \citenamefont {Burke},\ and\ \citenamefont {Ernzerhof}}]{perdew_1996}%
  \BibitemOpen
  \bibfield  {author} {\bibinfo {author} {\bibfnamefont {John~P.}\ \bibnamefont
  {Perdew}}, \bibinfo {author} {\bibfnamefont {Kieron}\ \bibnamefont {Burke}},
  \ and\ \bibinfo {author} {\bibfnamefont {Matthias}\ \bibnamefont
  {Ernzerhof}},\ }\bibfield  {title} {\enquote {\bibinfo {title} {Generalized
  gradient approximation made simple},}\ }\href {\doibase
  10.1103/PhysRevLett.77.3865} {\bibfield  {journal} {\bibinfo  {journal}
  {Phys. Rev. Lett.}\ }\textbf {\bibinfo {volume} {77}},\ \bibinfo {pages}
  {3865--3868} (\bibinfo {year} {1996})}\BibitemShut {NoStop}%
\bibitem [{\citenamefont {K{\"u}mmel}\ and\ \citenamefont
  {Kronik}(2008)}]{kummel_2008}%
  \BibitemOpen
  \bibfield  {author} {\bibinfo {author} {\bibfnamefont {Stephan}\ \bibnamefont
  {K{\"u}mmel}}\ and\ \bibinfo {author} {\bibfnamefont {Leeor}\ \bibnamefont
  {Kronik}},\ }\bibfield  {title} {\enquote {\bibinfo {title}
  {Orbital-dependent density functionals: Theory and applications},}\ }\href
  {\doibase 10.1103/RevModPhys.80.3} {\bibfield  {journal} {\bibinfo  {journal}
  {Rev. Mod. Phys.}\ }\textbf {\bibinfo {volume} {80}},\ \bibinfo {pages}
  {3--60} (\bibinfo {year} {2008})}\BibitemShut {NoStop}%
\bibitem [{\citenamefont {Skone}\ \emph {et~al.}(2014)\citenamefont {Skone},
  \citenamefont {Govoni},\ and\ \citenamefont {Galli}}]{skone_2014}%
  \BibitemOpen
  \bibfield  {author} {\bibinfo {author} {\bibfnamefont {Jonathan~H.}\
  \bibnamefont {Skone}}, \bibinfo {author} {\bibfnamefont {Marco}\ \bibnamefont
  {Govoni}}, \ and\ \bibinfo {author} {\bibfnamefont {Giulia}\ \bibnamefont
  {Galli}},\ }\bibfield  {title} {\enquote {\bibinfo {title} {Self-consistent
  hybrid functional for condensed systems},}\ }\href {\doibase
  10.1103/PhysRevB.89.195112} {\bibfield  {journal} {\bibinfo  {journal} {Phys.
  Rev. B}\ }\textbf {\bibinfo {volume} {89}},\ \bibinfo {pages} {195112}
  (\bibinfo {year} {2014})}\BibitemShut {NoStop}%
\bibitem [{\citenamefont {Heyd}\ \emph {et~al.}(2003)\citenamefont {Heyd},
  \citenamefont {Scuseria},\ and\ \citenamefont {Ernzerhof}}]{heyd_2003}%
  \BibitemOpen
  \bibfield  {author} {\bibinfo {author} {\bibfnamefont {Jochen}\ \bibnamefont
  {Heyd}}, \bibinfo {author} {\bibfnamefont {Gustavo~E.}\ \bibnamefont
  {Scuseria}}, \ and\ \bibinfo {author} {\bibfnamefont {Matthias}\ \bibnamefont
  {Ernzerhof}},\ }\bibfield  {title} {\enquote {\bibinfo {title} {Hybrid
  functionals based on a screened {Coulomb} potential},}\ }\href {\doibase
  10.1063/1.1564060} {\bibfield  {journal} {\bibinfo  {journal} {J. Chem.
  Phys.}\ }\textbf {\bibinfo {volume} {118}},\ \bibinfo {pages} {8207--8215}
  (\bibinfo {year} {2003})}\BibitemShut {NoStop}%
\bibitem [{\citenamefont {Heyd}\ \emph {et~al.}(2006)\citenamefont {Heyd},
  \citenamefont {Scuseria},\ and\ \citenamefont {Ernzerhof}}]{heyd_2006}%
  \BibitemOpen
  \bibfield  {author} {\bibinfo {author} {\bibfnamefont {Jochen}\ \bibnamefont
  {Heyd}}, \bibinfo {author} {\bibfnamefont {Gustavo~E.}\ \bibnamefont
  {Scuseria}}, \ and\ \bibinfo {author} {\bibfnamefont {Matthias}\ \bibnamefont
  {Ernzerhof}},\ }\bibfield  {title} {\enquote {\bibinfo {title} {Erratum:
  ``{Hybrid} functionals based on a screened {Coulomb} potential'' [{J.}
  {Chem.} {Phys.} 118, 8207 (2003)]},}\ }\href {\doibase 10.1063/1.2204597}
  {\bibfield  {journal} {\bibinfo  {journal} {J. Chem. Phys.}\ }\textbf
  {\bibinfo {volume} {124}},\ \bibinfo {pages} {219906} (\bibinfo {year}
  {2006})}\BibitemShut {NoStop}%
\bibitem [{\citenamefont {Hedin}(1965)}]{hedin_1965}%
  \BibitemOpen
  \bibfield  {author} {\bibinfo {author} {\bibfnamefont {Lars}\ \bibnamefont
  {Hedin}},\ }\bibfield  {title} {\enquote {\bibinfo {title} {New method for
  calculating the one-particle {Green's} function with application to the
  electron-gas problem},}\ }\href {\doibase 10.1103/PhysRev.139.A796}
  {\bibfield  {journal} {\bibinfo  {journal} {Phys. Rev.}\ }\textbf {\bibinfo
  {volume} {139}},\ \bibinfo {pages} {A796--A823} (\bibinfo {year}
  {1965})}\BibitemShut {NoStop}%
\bibitem [{\citenamefont {Hybertsen}\ and\ \citenamefont
  {Louie}(1986)}]{hybertsen_1986}%
  \BibitemOpen
  \bibfield  {author} {\bibinfo {author} {\bibfnamefont {Mark~S.}\ \bibnamefont
  {Hybertsen}}\ and\ \bibinfo {author} {\bibfnamefont {Steven~G.}\ \bibnamefont
  {Louie}},\ }\bibfield  {title} {\enquote {\bibinfo {title} {Electron
  correlation in semiconductors and insulators: Band gaps and quasiparticle
  energies},}\ }\href {\doibase 10.1103/PhysRevB.34.5390} {\bibfield  {journal}
  {\bibinfo  {journal} {Phys. Rev. B}\ }\textbf {\bibinfo {volume} {34}},\
  \bibinfo {pages} {5390--5413} (\bibinfo {year} {1986})}\BibitemShut {NoStop}%
\bibitem [{\citenamefont {Aryasetiawan}\ and\ \citenamefont
  {Gunnarsson}(1998)}]{aryasetiawan_1998}%
  \BibitemOpen
  \bibfield  {author} {\bibinfo {author} {\bibfnamefont {F.}~\bibnamefont
  {Aryasetiawan}}\ and\ \bibinfo {author} {\bibfnamefont {O.}~\bibnamefont
  {Gunnarsson}},\ }\bibfield  {title} {\enquote {\bibinfo {title} {The {GW}
  method},}\ }\href {\doibase 10.1088/0034-4885/61/3/002} {\bibfield  {journal}
  {\bibinfo  {journal} {Rep. Prog. Phys.}\ }\textbf {\bibinfo {volume} {61}},\
  \bibinfo {pages} {237} (\bibinfo {year} {1998})}\BibitemShut {NoStop}%
\bibitem [{\citenamefont {Govoni}\ and\ \citenamefont
  {Galli}(2015)}]{govoni_2015}%
  \BibitemOpen
  \bibfield  {author} {\bibinfo {author} {\bibfnamefont {Marco}\ \bibnamefont
  {Govoni}}\ and\ \bibinfo {author} {\bibfnamefont {Giulia}\ \bibnamefont
  {Galli}},\ }\bibfield  {title} {\enquote {\bibinfo {title} {Large scale {GW}
  calculations},}\ }\href {\doibase 10.1021/ct500958p} {\bibfield  {journal}
  {\bibinfo  {journal} {J. Chem. Theory Comput.}\ }\textbf {\bibinfo {volume}
  {11}},\ \bibinfo {pages} {2680--2696} (\bibinfo {year} {2015})}\BibitemShut
  {NoStop}%
\bibitem [{\citenamefont {Scherpelz}\ \emph {et~al.}(2016)\citenamefont
  {Scherpelz}, \citenamefont {Govoni}, \citenamefont {Hamada},\ and\
  \citenamefont {Galli}}]{scherpelz_2016}%
  \BibitemOpen
  \bibfield  {author} {\bibinfo {author} {\bibfnamefont {Peter}\ \bibnamefont
  {Scherpelz}}, \bibinfo {author} {\bibfnamefont {Marco}\ \bibnamefont
  {Govoni}}, \bibinfo {author} {\bibfnamefont {Ikutaro}\ \bibnamefont
  {Hamada}}, \ and\ \bibinfo {author} {\bibfnamefont {Giulia}\ \bibnamefont
  {Galli}},\ }\bibfield  {title} {\enquote {\bibinfo {title} {Implementation
  and validation of fully relativistic {GW} calculations: Spin-orbit coupling
  in molecules, nanocrystals, and solids},}\ }\href {\doibase
  10.1021/acs.jctc.6b00114} {\bibfield  {journal} {\bibinfo  {journal} {J.
  Chem. Theory Comput.}\ }\textbf {\bibinfo {volume} {12}},\ \bibinfo {pages}
  {3523--3544} (\bibinfo {year} {2016})}\BibitemShut {NoStop}%
\bibitem [{wes()}]{west_code}%
  \BibitemOpen
  \href {http://www.west-code.org/} {\enquote {\bibinfo {title} {{WEST}},}\
  }\bibinfo {note} {{http://www.west-code.org}, (accessed Jan 26,
  2016)}\BibitemShut {NoStop}%
\bibitem [{\citenamefont {Wilson}\ \emph {et~al.}(2008)\citenamefont {Wilson},
  \citenamefont {Gygi},\ and\ \citenamefont {Galli}}]{wilson_2008}%
  \BibitemOpen
  \bibfield  {author} {\bibinfo {author} {\bibfnamefont {Hugh~F.}\ \bibnamefont
  {Wilson}}, \bibinfo {author} {\bibfnamefont {Fran{\c{c}}ois}\ \bibnamefont
  {Gygi}}, \ and\ \bibinfo {author} {\bibfnamefont {Giulia}\ \bibnamefont
  {Galli}},\ }\bibfield  {title} {\enquote {\bibinfo {title} {Efficient
  iterative method for calculations of dielectric matrices},}\ }\href {\doibase
  10.1103/PhysRevB.78.113303} {\bibfield  {journal} {\bibinfo  {journal} {Phys.
  Rev. B}\ }\textbf {\bibinfo {volume} {78}},\ \bibinfo {pages} {113303}
  (\bibinfo {year} {2008})}\BibitemShut {NoStop}%
\bibitem [{\citenamefont {Wilson}\ \emph {et~al.}(2009)\citenamefont {Wilson},
  \citenamefont {Lu}, \citenamefont {Gygi},\ and\ \citenamefont
  {Galli}}]{wilson_2009}%
  \BibitemOpen
  \bibfield  {author} {\bibinfo {author} {\bibfnamefont {Hugh~F.}\ \bibnamefont
  {Wilson}}, \bibinfo {author} {\bibfnamefont {Deyu}\ \bibnamefont {Lu}},
  \bibinfo {author} {\bibfnamefont {Fran{\c{c}}ois}\ \bibnamefont {Gygi}}, \
  and\ \bibinfo {author} {\bibfnamefont {Giulia}\ \bibnamefont {Galli}},\
  }\bibfield  {title} {\enquote {\bibinfo {title} {Iterative calculations of
  dielectric eigenvalue spectra},}\ }\href {\doibase
  10.1103/PhysRevB.79.245106} {\bibfield  {journal} {\bibinfo  {journal} {Phys.
  Rev. B}\ }\textbf {\bibinfo {volume} {79}},\ \bibinfo {pages} {245106}
  (\bibinfo {year} {2009})}\BibitemShut {NoStop}%
\bibitem [{\citenamefont {Nguyen}\ \emph {et~al.}(2012)\citenamefont {Nguyen},
  \citenamefont {Pham}, \citenamefont {Rocca},\ and\ \citenamefont
  {Galli}}]{nguyen_2012}%
  \BibitemOpen
  \bibfield  {author} {\bibinfo {author} {\bibfnamefont {Huy-Viet}\
  \bibnamefont {Nguyen}}, \bibinfo {author} {\bibfnamefont {T.~Anh}\
  \bibnamefont {Pham}}, \bibinfo {author} {\bibfnamefont {Dario}\ \bibnamefont
  {Rocca}}, \ and\ \bibinfo {author} {\bibfnamefont {Giulia}\ \bibnamefont
  {Galli}},\ }\bibfield  {title} {\enquote {\bibinfo {title} {Improving
  accuracy and efficiency of calculations of photoemission spectra within the
  many-body perturbation theory},}\ }\href {\doibase
  10.1103/PhysRevB.85.081101} {\bibfield  {journal} {\bibinfo  {journal} {Phys.
  Rev. B}\ }\textbf {\bibinfo {volume} {85}},\ \bibinfo {pages} {081101}
  (\bibinfo {year} {2012})}\BibitemShut {NoStop}%
\bibitem [{\citenamefont {Pham}\ \emph {et~al.}(2013)\citenamefont {Pham},
  \citenamefont {Nguyen}, \citenamefont {Rocca},\ and\ \citenamefont
  {Galli}}]{pham_2013}%
  \BibitemOpen
  \bibfield  {author} {\bibinfo {author} {\bibfnamefont {T.~Anh}\ \bibnamefont
  {Pham}}, \bibinfo {author} {\bibfnamefont {Huy-Viet}\ \bibnamefont {Nguyen}},
  \bibinfo {author} {\bibfnamefont {Dario}\ \bibnamefont {Rocca}}, \ and\
  \bibinfo {author} {\bibfnamefont {Giulia}\ \bibnamefont {Galli}},\ }\bibfield
   {title} {\enquote {\bibinfo {title} {${GW}$ calculations using the spectral
  decomposition of the dielectric matrix: Verification, validation, and
  comparison of methods},}\ }\href {\doibase 10.1103/PhysRevB.87.155148}
  {\bibfield  {journal} {\bibinfo  {journal} {Phys. Rev. B}\ }\textbf {\bibinfo
  {volume} {87}},\ \bibinfo {pages} {155148} (\bibinfo {year}
  {2013})}\BibitemShut {NoStop}%
\bibitem [{\citenamefont {Baroni}\ \emph {et~al.}(2001)\citenamefont {Baroni},
  \citenamefont {de~Gironcoli}, \citenamefont {Dal~Corso},\ and\ \citenamefont
  {Giannozzi}}]{baroni_2001}%
  \BibitemOpen
  \bibfield  {author} {\bibinfo {author} {\bibfnamefont {Stefano}\ \bibnamefont
  {Baroni}}, \bibinfo {author} {\bibfnamefont {Stefano}\ \bibnamefont
  {de~Gironcoli}}, \bibinfo {author} {\bibfnamefont {Andrea}\ \bibnamefont
  {Dal~Corso}}, \ and\ \bibinfo {author} {\bibfnamefont {Paolo}\ \bibnamefont
  {Giannozzi}},\ }\bibfield  {title} {\enquote {\bibinfo {title} {Phonons and
  related crystal properties from density-functional perturbation theory},}\
  }\href {\doibase 10.1103/RevModPhys.73.515} {\bibfield  {journal} {\bibinfo
  {journal} {Rev. Mod. Phys.}\ }\textbf {\bibinfo {volume} {73}},\ \bibinfo
  {pages} {515--562} (\bibinfo {year} {2001})}\BibitemShut {NoStop}%
\bibitem [{\citenamefont {Wieferink}\ \emph {et~al.}(2010)\citenamefont
  {Wieferink}, \citenamefont {Kr{\"u}ger},\ and\ \citenamefont
  {Pollmann}}]{wieferink_2010}%
  \BibitemOpen
  \bibfield  {author} {\bibinfo {author} {\bibfnamefont {J{\"u}rgen}\
  \bibnamefont {Wieferink}}, \bibinfo {author} {\bibfnamefont {Peter}\
  \bibnamefont {Kr{\"u}ger}}, \ and\ \bibinfo {author} {\bibfnamefont
  {Johannes}\ \bibnamefont {Pollmann}},\ }\bibfield  {title} {\enquote
  {\bibinfo {title} {Ab initio study of atomic hydrogen diffusion on the clean
  and hydrogen-terminated {Si}(001) surface},}\ }\href {\doibase
  10.1103/PhysRevB.82.075323} {\bibfield  {journal} {\bibinfo  {journal} {Phys.
  Rev. B}\ }\textbf {\bibinfo {volume} {82}},\ \bibinfo {pages} {075323}
  (\bibinfo {year} {2010})}\BibitemShut {NoStop}%
\bibitem [{\citenamefont {Blomquist}\ and\ \citenamefont
  {Kirczenow}(2006)}]{blomquist_2006}%
  \BibitemOpen
  \bibfield  {author} {\bibinfo {author} {\bibfnamefont {Torbj{\"o}rn}\
  \bibnamefont {Blomquist}}\ and\ \bibinfo {author} {\bibfnamefont {George}\
  \bibnamefont {Kirczenow}},\ }\bibfield  {title} {\enquote {\bibinfo {title}
  {Controlling the charge of a specific surface atom by the addition of a
  non-site-specific single impurity in a {Si} nanocrystal},}\ }\href {\doibase
  10.1021/nl051995s} {\bibfield  {journal} {\bibinfo  {journal} {Nano Lett.}\
  }\textbf {\bibinfo {volume} {6}},\ \bibinfo {pages} {61--65} (\bibinfo {year}
  {2006})}\BibitemShut {NoStop}%
\bibitem [{\citenamefont {Haider}\ \emph {et~al.}(2008)\citenamefont {Haider},
  \citenamefont {Pitters}, \citenamefont {{DiLabio}}, \citenamefont {Livadaru},
  \citenamefont {Mutus},\ and\ \citenamefont {Wolkow}}]{haider_2008}%
  \BibitemOpen
  \bibfield  {author} {\bibinfo {author} {\bibfnamefont {M.~Baseer}\
  \bibnamefont {Haider}}, \bibinfo {author} {\bibfnamefont {Jason~L.}\
  \bibnamefont {Pitters}}, \bibinfo {author} {\bibfnamefont {Gino~A.}\
  \bibnamefont {{DiLabio}}}, \bibinfo {author} {\bibfnamefont {Lucian}\
  \bibnamefont {Livadaru}}, \bibinfo {author} {\bibfnamefont {Josh~Y.}\
  \bibnamefont {Mutus}}, \ and\ \bibinfo {author} {\bibfnamefont {Robert~A.}\
  \bibnamefont {Wolkow}},\ }\bibfield  {title} {\enquote {\bibinfo {title}
  {Controlled coupling and occupation of silicon atomic quantum dots},}\ }\href
  {http://arxiv.org/abs/0807.0609} {\  (\bibinfo {year} {2008})},\ \Eprint
  {http://arxiv.org/abs/0807.0609 [cond-mat]} {arXiv:0807.0609 [cond-mat]}
  \BibitemShut {NoStop}%
\bibitem [{\citenamefont {Raza}(2007)}]{raza_2007}%
  \BibitemOpen
  \bibfield  {author} {\bibinfo {author} {\bibfnamefont {Hassan}\ \bibnamefont
  {Raza}},\ }\bibfield  {title} {\enquote {\bibinfo {title} {Theoretical study
  of isolated dangling bonds, dangling bond wires, and dangling bond clusters
  on a {H:Si}(001)-2x1 surface},}\ }\href {\doibase 10.1103/PhysRevB.76.045308}
  {\bibfield  {journal} {\bibinfo  {journal} {Phys. Rev. B}\ }\textbf {\bibinfo
  {volume} {76}},\ \bibinfo {pages} {045308} (\bibinfo {year}
  {2007})}\BibitemShut {NoStop}%
\bibitem [{\citenamefont {Komsa}\ and\ \citenamefont
  {Pasquarello}(2013)}]{komsa_2013}%
  \BibitemOpen
  \bibfield  {author} {\bibinfo {author} {\bibfnamefont {Hannu-Pekka}\
  \bibnamefont {Komsa}}\ and\ \bibinfo {author} {\bibfnamefont {Alfredo}\
  \bibnamefont {Pasquarello}},\ }\bibfield  {title} {\enquote {\bibinfo {title}
  {Finite-size supercell correction for charged defects at surfaces and
  interfaces},}\ }\href {\doibase 10.1103/PhysRevLett.110.095505} {\bibfield
  {journal} {\bibinfo  {journal} {Phys. Rev. Lett.}\ }\textbf {\bibinfo
  {volume} {110}},\ \bibinfo {pages} {095505} (\bibinfo {year}
  {2013})}\BibitemShut {NoStop}%
\bibitem [{Note1()}]{Note1}%
  \BibitemOpen
  \bibinfo {note} {Only the lowest two values from Refs.\ \cite
  {wieferink_2010, schofield_2013} are explicitly spin-polarized; the
  calculations that cluster at 0.35-0.42 eV are thus unlikely to be physically
  accurate.}\BibitemShut {Stop}%
\bibitem [{\citenamefont {Li}\ and\ \citenamefont {Galli}(2010)}]{li_2010}%
  \BibitemOpen
  \bibfield  {author} {\bibinfo {author} {\bibfnamefont {Yan}\ \bibnamefont
  {Li}}\ and\ \bibinfo {author} {\bibfnamefont {Giulia}\ \bibnamefont
  {Galli}},\ }\bibfield  {title} {\enquote {\bibinfo {title} {Electronic and
  spectroscopic properties of the hydrogen-terminated {Si(111)} surface from ab
  initio calculations},}\ }\href {\doibase 10.1103/PhysRevB.82.045321}
  {\bibfield  {journal} {\bibinfo  {journal} {Phys. Rev. B}\ }\textbf {\bibinfo
  {volume} {82}},\ \bibinfo {pages} {045321} (\bibinfo {year}
  {2010})}\BibitemShut {NoStop}%
\bibitem [{\citenamefont {Fischetti}\ \emph {et~al.}(2011)\citenamefont
  {Fischetti}, \citenamefont {Fu}, \citenamefont {Narayanan},\ and\
  \citenamefont {Kim}}]{fischetti_2011}%
  \BibitemOpen
  \bibfield  {author} {\bibinfo {author} {\bibfnamefont {Massimo~V.}\
  \bibnamefont {Fischetti}}, \bibinfo {author} {\bibfnamefont {Bo}~\bibnamefont
  {Fu}}, \bibinfo {author} {\bibfnamefont {Sudarshan}\ \bibnamefont
  {Narayanan}}, \ and\ \bibinfo {author} {\bibfnamefont {Jiseok}\ \bibnamefont
  {Kim}},\ }\bibfield  {title} {\enquote {\bibinfo {title} {Semiclassical and
  quantum electronic transport in nanometer-scale structures: Empirical
  pseudopotential band structure, {Monte} {Carlo} simulations and {Pauli}
  master equation},}\ }in\ \href@noop {} {\emph {\bibinfo {booktitle}
  {Nano-Electronic Devices}}},\ \bibinfo {editor} {edited by\ \bibinfo {editor}
  {\bibfnamefont {Dragica}\ \bibnamefont {Vasileska}}\ and\ \bibinfo {editor}
  {\bibfnamefont {Stephen~M.}\ \bibnamefont {Goodnick}}}\ (\bibinfo
  {publisher} {Springer New York},\ \bibinfo {year} {2011})\ pp.\ \bibinfo
  {pages} {183--247}\BibitemShut {NoStop}%
\bibitem [{\citenamefont {Sagisaka}\ \emph {et~al.}(2017)\citenamefont
  {Sagisaka}, \citenamefont {Nara},\ and\ \citenamefont
  {Bowler}}]{sagisaka_2017}%
  \BibitemOpen
  \bibfield  {author} {\bibinfo {author} {\bibfnamefont {Keisuke}\ \bibnamefont
  {Sagisaka}}, \bibinfo {author} {\bibfnamefont {Jun}\ \bibnamefont {Nara}}, \
  and\ \bibinfo {author} {\bibfnamefont {David}\ \bibnamefont {Bowler}},\
  }\bibfield  {title} {\enquote {\bibinfo {title} {Importance of bulk states
  for the electronic structure of semiconductor surfaces: implications for
  finite slabs},}\ }\href {\doibase 10.1088/1361-648X/aa5f91} {\bibfield
  {journal} {\bibinfo  {journal} {J. Phys.: Condens. Matter}\ } (\bibinfo
  {year} {2017}),\ 10.1088/1361-648X/aa5f91}\BibitemShut {NoStop}%
\bibitem [{\citenamefont {Lenahan}\ and\ \citenamefont
  {Conley~Jr.}(1998)}]{lenahan_1998}%
  \BibitemOpen
  \bibfield  {author} {\bibinfo {author} {\bibfnamefont {P.~M.}\ \bibnamefont
  {Lenahan}}\ and\ \bibinfo {author} {\bibfnamefont {J.~F.}\ \bibnamefont
  {Conley~Jr.}},\ }\bibfield  {title} {\enquote {\bibinfo {title} {What can
  electron paramagnetic resonance tell us about the {Si}/{SiO}${}_2$ system?}}\
  }\href {\doibase 10.1116/1.590301} {\bibfield  {journal} {\bibinfo  {journal}
  {J. Vac. Sci. Technol. B}\ }\textbf {\bibinfo {volume} {16}},\ \bibinfo
  {pages} {2134} (\bibinfo {year} {1998})}\BibitemShut {NoStop}%
\bibitem [{\citenamefont {Freysoldt}\ \emph {et~al.}(2014)\citenamefont
  {Freysoldt}, \citenamefont {Grabowski}, \citenamefont {Hickel}, \citenamefont
  {Neugebauer}, \citenamefont {Kresse}, \citenamefont {Janotti},\ and\
  \citenamefont {Van~de Walle}}]{freysoldt_2014}%
  \BibitemOpen
  \bibfield  {author} {\bibinfo {author} {\bibfnamefont {Christoph}\
  \bibnamefont {Freysoldt}}, \bibinfo {author} {\bibfnamefont {Blazej}\
  \bibnamefont {Grabowski}}, \bibinfo {author} {\bibfnamefont {Tilmann}\
  \bibnamefont {Hickel}}, \bibinfo {author} {\bibfnamefont {J{\"o}rg}\
  \bibnamefont {Neugebauer}}, \bibinfo {author} {\bibfnamefont {Georg}\
  \bibnamefont {Kresse}}, \bibinfo {author} {\bibfnamefont {Anderson}\
  \bibnamefont {Janotti}}, \ and\ \bibinfo {author} {\bibfnamefont {Chris~G.}\
  \bibnamefont {Van~de Walle}},\ }\bibfield  {title} {\enquote {\bibinfo
  {title} {First-principles calculations for point defects in solids},}\ }\href
  {\doibase 10.1103/RevModPhys.86.253} {\bibfield  {journal} {\bibinfo
  {journal} {Rev. Mod. Phys.}\ }\textbf {\bibinfo {volume} {86}},\ \bibinfo
  {pages} {253--305} (\bibinfo {year} {2014})}\BibitemShut {NoStop}%
\bibitem [{\citenamefont {Shaterzadeh-Yazdi}\ \emph {et~al.}(2014)\citenamefont
  {Shaterzadeh-Yazdi}, \citenamefont {Livadaru}, \citenamefont {Taucer},
  \citenamefont {Mutus}, \citenamefont {Pitters}, \citenamefont {Wolkow},\ and\
  \citenamefont {Sanders}}]{shaterzadeh-yazdi_2014}%
  \BibitemOpen
  \bibfield  {author} {\bibinfo {author} {\bibfnamefont {Zahra}\ \bibnamefont
  {Shaterzadeh-Yazdi}}, \bibinfo {author} {\bibfnamefont {Lucian}\ \bibnamefont
  {Livadaru}}, \bibinfo {author} {\bibfnamefont {Marco}\ \bibnamefont
  {Taucer}}, \bibinfo {author} {\bibfnamefont {Josh}\ \bibnamefont {Mutus}},
  \bibinfo {author} {\bibfnamefont {Jason}\ \bibnamefont {Pitters}}, \bibinfo
  {author} {\bibfnamefont {Robert~A.}\ \bibnamefont {Wolkow}}, \ and\ \bibinfo
  {author} {\bibfnamefont {Barry~C.}\ \bibnamefont {Sanders}},\ }\bibfield
  {title} {\enquote {\bibinfo {title} {Characterizing the rate and coherence of
  single-electron tunneling between two dangling bonds on the surface of
  silicon},}\ }\href {\doibase 10.1103/PhysRevB.89.035315} {\bibfield
  {journal} {\bibinfo  {journal} {Phys. Rev. B}\ }\textbf {\bibinfo {volume}
  {89}},\ \bibinfo {pages} {035315} (\bibinfo {year} {2014})}\BibitemShut
  {NoStop}%
\bibitem [{\citenamefont {Lee}\ \emph {et~al.}(2008)\citenamefont {Lee},
  \citenamefont {Choi},\ and\ \citenamefont {Cho}}]{lee_2008}%
  \BibitemOpen
  \bibfield  {author} {\bibinfo {author} {\bibfnamefont {Ji~Young}\
  \bibnamefont {Lee}}, \bibinfo {author} {\bibfnamefont {Jin-Ho}\ \bibnamefont
  {Choi}}, \ and\ \bibinfo {author} {\bibfnamefont {Jun-Hyung}\ \bibnamefont
  {Cho}},\ }\bibfield  {title} {\enquote {\bibinfo {title} {Antiferromagnetic
  coupling between two adjacent dangling bonds on si(001): Total-energy and
  force calculations},}\ }\href {\doibase 10.1103/PhysRevB.78.081303}
  {\bibfield  {journal} {\bibinfo  {journal} {Phys. Rev. B}\ }\textbf {\bibinfo
  {volume} {78}},\ \bibinfo {pages} {081303} (\bibinfo {year}
  {2008})}\BibitemShut {NoStop}%
\bibitem [{\citenamefont {Lee}\ \emph {et~al.}(2009)\citenamefont {Lee},
  \citenamefont {Cho},\ and\ \citenamefont {Zhang}}]{lee_2009}%
  \BibitemOpen
  \bibfield  {author} {\bibinfo {author} {\bibfnamefont {Ji~Young}\
  \bibnamefont {Lee}}, \bibinfo {author} {\bibfnamefont {Jun-Hyung}\
  \bibnamefont {Cho}}, \ and\ \bibinfo {author} {\bibfnamefont {Zhenyu}\
  \bibnamefont {Zhang}},\ }\bibfield  {title} {\enquote {\bibinfo {title}
  {Quantum size effects in competing charge and spin orderings of dangling bond
  wires on {Si}(001)},}\ }\href {\doibase 10.1103/PhysRevB.80.155329}
  {\bibfield  {journal} {\bibinfo  {journal} {Phys. Rev. B}\ }\textbf {\bibinfo
  {volume} {80}},\ \bibinfo {pages} {155329} (\bibinfo {year}
  {2009})}\BibitemShut {NoStop}%
\bibitem [{\citenamefont {Robles}\ \emph {et~al.}(2012)\citenamefont {Robles},
  \citenamefont {Kepenekian}, \citenamefont {Monturet}, \citenamefont
  {Joachim},\ and\ \citenamefont {Lorente}}]{robles_2012}%
  \BibitemOpen
  \bibfield  {author} {\bibinfo {author} {\bibfnamefont {R.}~\bibnamefont
  {Robles}}, \bibinfo {author} {\bibfnamefont {M.}~\bibnamefont {Kepenekian}},
  \bibinfo {author} {\bibfnamefont {S.}~\bibnamefont {Monturet}}, \bibinfo
  {author} {\bibfnamefont {C.}~\bibnamefont {Joachim}}, \ and\ \bibinfo
  {author} {\bibfnamefont {N.}~\bibnamefont {Lorente}},\ }\bibfield  {title}
  {\enquote {\bibinfo {title} {Energetics and stability of dangling-bond
  silicon wires on {H} passivated {Si}(100)},}\ }\href {\doibase
  10.1088/0953-8984/24/44/445004} {\bibfield  {journal} {\bibinfo  {journal}
  {J. Phys.: Condens. Matter}\ }\textbf {\bibinfo {volume} {24}},\ \bibinfo
  {pages} {445004} (\bibinfo {year} {2012})}\BibitemShut {NoStop}%
\bibitem [{\citenamefont {Kepenekian}\ \emph {et~al.}(2013)\citenamefont
  {Kepenekian}, \citenamefont {Novaes}, \citenamefont {Robles}, \citenamefont
  {Monturet}, \citenamefont {Kawai}, \citenamefont {{Christian Joachim}},\ and\
  \citenamefont {Lorente}}]{kepenekian_2013}%
  \BibitemOpen
  \bibfield  {author} {\bibinfo {author} {\bibfnamefont {Mika{\"e}l}\
  \bibnamefont {Kepenekian}}, \bibinfo {author} {\bibfnamefont {Frederico~D.}\
  \bibnamefont {Novaes}}, \bibinfo {author} {\bibfnamefont {Roberto}\
  \bibnamefont {Robles}}, \bibinfo {author} {\bibfnamefont {Serge}\
  \bibnamefont {Monturet}}, \bibinfo {author} {\bibfnamefont {Hiroyo}\
  \bibnamefont {Kawai}}, \bibinfo {author} {\bibnamefont {{Christian
  Joachim}}}, \ and\ \bibinfo {author} {\bibfnamefont {Nicol{\'a}s}\
  \bibnamefont {Lorente}},\ }\bibfield  {title} {\enquote {\bibinfo {title}
  {Electron transport through dangling-bond silicon wires on {H}-passivated
  {Si}(100)},}\ }\href {\doibase 10.1088/0953-8984/25/2/025503} {\bibfield
  {journal} {\bibinfo  {journal} {J. Phys.: Condens. Matter}\ }\textbf
  {\bibinfo {volume} {25}},\ \bibinfo {pages} {025503} (\bibinfo {year}
  {2013})}\BibitemShut {NoStop}%
\bibitem [{\citenamefont {Engelund}\ \emph {et~al.}(2016)\citenamefont
  {Engelund}, \citenamefont {Papior}, \citenamefont {Brandimarte},
  \citenamefont {Frederiksen}, \citenamefont {Garcia-Lekue},\ and\
  \citenamefont {S{\'a}nchez-Portal}}]{engelund_2016}%
  \BibitemOpen
  \bibfield  {author} {\bibinfo {author} {\bibfnamefont {Mads}\ \bibnamefont
  {Engelund}}, \bibinfo {author} {\bibfnamefont {Nick}\ \bibnamefont {Papior}},
  \bibinfo {author} {\bibfnamefont {Pedro}\ \bibnamefont {Brandimarte}},
  \bibinfo {author} {\bibfnamefont {Thomas}\ \bibnamefont {Frederiksen}},
  \bibinfo {author} {\bibfnamefont {Aran}\ \bibnamefont {Garcia-Lekue}}, \ and\
  \bibinfo {author} {\bibfnamefont {Daniel}\ \bibnamefont
  {S{\'a}nchez-Portal}},\ }\bibfield  {title} {\enquote {\bibinfo {title}
  {Search for a metallic dangling-bond wire on {n}-doped {H}-passivated
  semiconductor surfaces},}\ }\href {\doibase 10.1021/acs.jpcc.6b04540}
  {\bibfield  {journal} {\bibinfo  {journal} {J. Phys. Chem. C}\ }\textbf
  {\bibinfo {volume} {120}},\ \bibinfo {pages} {20303--20309} (\bibinfo {year}
  {2016})}\BibitemShut {NoStop}%
\bibitem [{\citenamefont {Zhou}\ \emph {et~al.}(2013)\citenamefont {Zhou},
  \citenamefont {Liu}, \citenamefont {Wang}, \citenamefont {Bai}, \citenamefont
  {Feng}, \citenamefont {Lagally},\ and\ \citenamefont {Liu}}]{zhou_2013}%
  \BibitemOpen
  \bibfield  {author} {\bibinfo {author} {\bibfnamefont {Miao}\ \bibnamefont
  {Zhou}}, \bibinfo {author} {\bibfnamefont {Zheng}\ \bibnamefont {Liu}},
  \bibinfo {author} {\bibfnamefont {Zhengfei}\ \bibnamefont {Wang}}, \bibinfo
  {author} {\bibfnamefont {Zhaoqiang}\ \bibnamefont {Bai}}, \bibinfo {author}
  {\bibfnamefont {Yuanping}\ \bibnamefont {Feng}}, \bibinfo {author}
  {\bibfnamefont {Max~G.}\ \bibnamefont {Lagally}}, \ and\ \bibinfo {author}
  {\bibfnamefont {Feng}\ \bibnamefont {Liu}},\ }\bibfield  {title} {\enquote
  {\bibinfo {title} {Strain-engineered surface transport in {Si(001)}: Complete
  isolation of the surface state via tensile strain},}\ }\href {\doibase
  10.1103/PhysRevLett.111.246801} {\bibfield  {journal} {\bibinfo  {journal}
  {Phys. Rev. Lett.}\ }\textbf {\bibinfo {volume} {111}},\ \bibinfo {pages}
  {246801} (\bibinfo {year} {2013})}\BibitemShut {NoStop}%
\bibitem [{\citenamefont {Mungu{\'{i}}a}\ \emph {et~al.}(2008)\citenamefont
  {Mungu{\'{i}}a}, \citenamefont {Bremond}, \citenamefont {Bluet},
  \citenamefont {Hartmann},\ and\ \citenamefont {Mermoux}}]{munguia_2008}%
  \BibitemOpen
  \bibfield  {author} {\bibinfo {author} {\bibfnamefont {J.}~\bibnamefont
  {Mungu{\'{i}}a}}, \bibinfo {author} {\bibfnamefont {G.}~\bibnamefont
  {Bremond}}, \bibinfo {author} {\bibfnamefont {J.~M.}\ \bibnamefont {Bluet}},
  \bibinfo {author} {\bibfnamefont {J.~M.}\ \bibnamefont {Hartmann}}, \ and\
  \bibinfo {author} {\bibfnamefont {M.}~\bibnamefont {Mermoux}},\ }\bibfield
  {title} {\enquote {\bibinfo {title} {Strain dependence of indirect band gap
  for strained silicon on insulator wafers},}\ }\href {\doibase
  10.1063/1.2978241} {\bibfield  {journal} {\bibinfo  {journal} {Appl. Phys.
  Lett.}\ }\textbf {\bibinfo {volume} {93}},\ \bibinfo {pages} {102101}
  (\bibinfo {year} {2008})}\BibitemShut {NoStop}%
\bibitem [{\citenamefont {Richard}\ \emph {et~al.}(2003)\citenamefont
  {Richard}, \citenamefont {Aniel}, \citenamefont {Fishman},\ and\
  \citenamefont {Cavassilas}}]{richard_2003}%
  \BibitemOpen
  \bibfield  {author} {\bibinfo {author} {\bibfnamefont {Soline}\ \bibnamefont
  {Richard}}, \bibinfo {author} {\bibfnamefont {Fr{\'{e}}d{\'{e}}ric}\
  \bibnamefont {Aniel}}, \bibinfo {author} {\bibfnamefont {Guy}\ \bibnamefont
  {Fishman}}, \ and\ \bibinfo {author} {\bibfnamefont {Nicolas}\ \bibnamefont
  {Cavassilas}},\ }\bibfield  {title} {\enquote {\bibinfo {title} {Energy-band
  structure in strained silicon: A 20-band k$\cdot$p and {Bir-Pikus}
  {Hamiltonian} model},}\ }\href {\doibase 10.1063/1.1587004} {\bibfield
  {journal} {\bibinfo  {journal} {J. Appl. Phys.}\ }\textbf {\bibinfo {volume}
  {94}},\ \bibinfo {pages} {1795--1799} (\bibinfo {year} {2003})}\BibitemShut
  {NoStop}%
\bibitem [{\citenamefont {Liu}\ \emph {et~al.}(2002)\citenamefont {Liu},
  \citenamefont {Huang}, \citenamefont {Rugheimer}, \citenamefont {Savage},\
  and\ \citenamefont {Lagally}}]{liu_2002}%
  \BibitemOpen
  \bibfield  {author} {\bibinfo {author} {\bibfnamefont {Feng}\ \bibnamefont
  {Liu}}, \bibinfo {author} {\bibfnamefont {Minghuang}\ \bibnamefont {Huang}},
  \bibinfo {author} {\bibfnamefont {P.~P.}\ \bibnamefont {Rugheimer}}, \bibinfo
  {author} {\bibfnamefont {D.~E.}\ \bibnamefont {Savage}}, \ and\ \bibinfo
  {author} {\bibfnamefont {M.~G.}\ \bibnamefont {Lagally}},\ }\bibfield
  {title} {\enquote {\bibinfo {title} {Nanostressors and the nanomechanical
  response of a thin silicon film on an insulator},}\ }\href {\doibase
  10.1103/PhysRevLett.89.136101} {\bibfield  {journal} {\bibinfo  {journal}
  {Phys. Rev. Lett.}\ }\textbf {\bibinfo {volume} {89}},\ \bibinfo {pages}
  {136101} (\bibinfo {year} {2002})}\BibitemShut {NoStop}%
\bibitem [{\citenamefont {Mungu{\'{i}}a}\ \emph {et~al.}(2012)\citenamefont
  {Mungu{\'{i}}a}, \citenamefont {Bluet}, \citenamefont {Marty}, \citenamefont
  {Bremond}, \citenamefont {Mermoux},\ and\ \citenamefont
  {Rouchon}}]{munguia_2012}%
  \BibitemOpen
  \bibfield  {author} {\bibinfo {author} {\bibfnamefont {J.}~\bibnamefont
  {Mungu{\'{i}}a}}, \bibinfo {author} {\bibfnamefont {J.-M.}\ \bibnamefont
  {Bluet}}, \bibinfo {author} {\bibfnamefont {O.}~\bibnamefont {Marty}},
  \bibinfo {author} {\bibfnamefont {G.}~\bibnamefont {Bremond}}, \bibinfo
  {author} {\bibfnamefont {M.}~\bibnamefont {Mermoux}}, \ and\ \bibinfo
  {author} {\bibfnamefont {D.}~\bibnamefont {Rouchon}},\ }\bibfield  {title}
  {\enquote {\bibinfo {title} {Temperature dependence of the indirect bandgap
  in ultrathin strained silicon on insulator layer},}\ }\href {\doibase
  10.1063/1.3691955} {\bibfield  {journal} {\bibinfo  {journal} {Appl. Phys.
  Lett.}\ }\textbf {\bibinfo {volume} {100}},\ \bibinfo {pages} {102107}
  (\bibinfo {year} {2012})}\BibitemShut {NoStop}%
\bibitem [{\citenamefont {Saeedi}\ \emph {et~al.}(2013)\citenamefont {Saeedi},
  \citenamefont {Simmons}, \citenamefont {Salvail}, \citenamefont {Dluhy},
  \citenamefont {Riemann}, \citenamefont {Abrosimov}, \citenamefont {Becker},
  \citenamefont {Pohl}, \citenamefont {Morton},\ and\ \citenamefont
  {Thewalt}}]{saeedi_2013}%
  \BibitemOpen
  \bibfield  {author} {\bibinfo {author} {\bibfnamefont {Kamyar}\ \bibnamefont
  {Saeedi}}, \bibinfo {author} {\bibfnamefont {Stephanie}\ \bibnamefont
  {Simmons}}, \bibinfo {author} {\bibfnamefont {Jeff~Z.}\ \bibnamefont
  {Salvail}}, \bibinfo {author} {\bibfnamefont {Phillip}\ \bibnamefont
  {Dluhy}}, \bibinfo {author} {\bibfnamefont {Helge}\ \bibnamefont {Riemann}},
  \bibinfo {author} {\bibfnamefont {Nikolai~V.}\ \bibnamefont {Abrosimov}},
  \bibinfo {author} {\bibfnamefont {Peter}\ \bibnamefont {Becker}}, \bibinfo
  {author} {\bibfnamefont {Hans-Joachim}\ \bibnamefont {Pohl}}, \bibinfo
  {author} {\bibfnamefont {John J.~L.}\ \bibnamefont {Morton}}, \ and\ \bibinfo
  {author} {\bibfnamefont {Mike L.~W.}\ \bibnamefont {Thewalt}},\ }\bibfield
  {title} {\enquote {\bibinfo {title} {Room-temperature quantum bit storage
  exceeding 39 minutes using ionized donors in silicon-28},}\ }\href {\doibase
  10.1126/science.1239584} {\bibfield  {journal} {\bibinfo  {journal}
  {Science}\ }\textbf {\bibinfo {volume} {342}},\ \bibinfo {pages} {830--833}
  (\bibinfo {year} {2013})}\BibitemShut {NoStop}%
\bibitem [{\citenamefont {Pla}\ \emph {et~al.}(2013)\citenamefont {Pla},
  \citenamefont {Tan}, \citenamefont {Dehollain}, \citenamefont {Lim},
  \citenamefont {Morton}, \citenamefont {Zwanenburg}, \citenamefont {Jamieson},
  \citenamefont {Dzurak},\ and\ \citenamefont {Morello}}]{pla_2013}%
  \BibitemOpen
  \bibfield  {author} {\bibinfo {author} {\bibfnamefont {Jarryd~J.}\
  \bibnamefont {Pla}}, \bibinfo {author} {\bibfnamefont {Kuan~Y.}\ \bibnamefont
  {Tan}}, \bibinfo {author} {\bibfnamefont {Juan~P.}\ \bibnamefont
  {Dehollain}}, \bibinfo {author} {\bibfnamefont {Wee~H.}\ \bibnamefont {Lim}},
  \bibinfo {author} {\bibfnamefont {John J.~L.}\ \bibnamefont {Morton}},
  \bibinfo {author} {\bibfnamefont {Floris~A.}\ \bibnamefont {Zwanenburg}},
  \bibinfo {author} {\bibfnamefont {David~N.}\ \bibnamefont {Jamieson}},
  \bibinfo {author} {\bibfnamefont {Andrew~S.}\ \bibnamefont {Dzurak}}, \ and\
  \bibinfo {author} {\bibfnamefont {Andrea}\ \bibnamefont {Morello}},\
  }\bibfield  {title} {\enquote {\bibinfo {title} {High-fidelity readout and
  control of a nuclear spin qubit in silicon},}\ }\href {\doibase
  10.1038/nature12011} {\bibfield  {journal} {\bibinfo  {journal} {Nature}\
  }\textbf {\bibinfo {volume} {496}},\ \bibinfo {pages} {334--338} (\bibinfo
  {year} {2013})}\BibitemShut {NoStop}%
\bibitem [{\citenamefont {Salfi}\ \emph {et~al.}(2016)\citenamefont {Salfi},
  \citenamefont {Mol}, \citenamefont {Rahman}, \citenamefont {Klimeck},
  \citenamefont {Simmons}, \citenamefont {Hollenberg},\ and\ \citenamefont
  {Rogge}}]{salfi_2016}%
  \BibitemOpen
  \bibfield  {author} {\bibinfo {author} {\bibfnamefont {J.}~\bibnamefont
  {Salfi}}, \bibinfo {author} {\bibfnamefont {J.~A.}\ \bibnamefont {Mol}},
  \bibinfo {author} {\bibfnamefont {R.}~\bibnamefont {Rahman}}, \bibinfo
  {author} {\bibfnamefont {G.}~\bibnamefont {Klimeck}}, \bibinfo {author}
  {\bibfnamefont {M.~Y.}\ \bibnamefont {Simmons}}, \bibinfo {author}
  {\bibfnamefont {L.~C.~L.}\ \bibnamefont {Hollenberg}}, \ and\ \bibinfo
  {author} {\bibfnamefont {S.}~\bibnamefont {Rogge}},\ }\bibfield  {title}
  {\enquote {\bibinfo {title} {Quantum simulation of the {Hubbard} model with
  dopant atoms in silicon},}\ }\href {\doibase 10.1038/ncomms11342} {\bibfield
  {journal} {\bibinfo  {journal} {Nat. Commun.}\ }\textbf {\bibinfo {volume}
  {7}},\ \bibinfo {pages} {11342} (\bibinfo {year} {2016})}\BibitemShut
  {NoStop}%
\bibitem [{\citenamefont {Mohammady}\ \emph {et~al.}(2012)\citenamefont
  {Mohammady}, \citenamefont {Morley}, \citenamefont {Nazir},\ and\
  \citenamefont {Monteiro}}]{mohammady_2012}%
  \BibitemOpen
  \bibfield  {author} {\bibinfo {author} {\bibfnamefont {M.~H.}\ \bibnamefont
  {Mohammady}}, \bibinfo {author} {\bibfnamefont {G.~W.}\ \bibnamefont
  {Morley}}, \bibinfo {author} {\bibfnamefont {A.}~\bibnamefont {Nazir}}, \
  and\ \bibinfo {author} {\bibfnamefont {T.~S.}\ \bibnamefont {Monteiro}},\
  }\bibfield  {title} {\enquote {\bibinfo {title} {Analysis of quantum
  coherence in bismuth-doped silicon: A system of strongly coupled spin
  qubits},}\ }\href {\doibase 10.1103/PhysRevB.85.094404} {\bibfield  {journal}
  {\bibinfo  {journal} {Phys. Rev. B}\ }\textbf {\bibinfo {volume} {85}},\
  \bibinfo {pages} {094404} (\bibinfo {year} {2012})}\BibitemShut {NoStop}%
\bibitem [{\citenamefont {Wolfowicz}\ \emph {et~al.}(2013)\citenamefont
  {Wolfowicz}, \citenamefont {Tyryshkin}, \citenamefont {George}, \citenamefont
  {Riemann}, \citenamefont {Abrosimov}, \citenamefont {Becker}, \citenamefont
  {Pohl}, \citenamefont {Thewalt}, \citenamefont {Lyon},\ and\ \citenamefont
  {Morton}}]{wolfowicz_2013}%
  \BibitemOpen
  \bibfield  {author} {\bibinfo {author} {\bibfnamefont {Gary}\ \bibnamefont
  {Wolfowicz}}, \bibinfo {author} {\bibfnamefont {Alexei~M.}\ \bibnamefont
  {Tyryshkin}}, \bibinfo {author} {\bibfnamefont {Richard~E.}\ \bibnamefont
  {George}}, \bibinfo {author} {\bibfnamefont {Helge}\ \bibnamefont {Riemann}},
  \bibinfo {author} {\bibfnamefont {Nikolai~V.}\ \bibnamefont {Abrosimov}},
  \bibinfo {author} {\bibfnamefont {Peter}\ \bibnamefont {Becker}}, \bibinfo
  {author} {\bibfnamefont {Hans-Joachim}\ \bibnamefont {Pohl}}, \bibinfo
  {author} {\bibfnamefont {Mike L.~W.}\ \bibnamefont {Thewalt}}, \bibinfo
  {author} {\bibfnamefont {Stephen~A.}\ \bibnamefont {Lyon}}, \ and\ \bibinfo
  {author} {\bibfnamefont {John J.~L.}\ \bibnamefont {Morton}},\ }\bibfield
  {title} {\enquote {\bibinfo {title} {Atomic clock transitions in
  silicon-based spin qubits},}\ }\href {\doibase 10.1038/nnano.2013.117}
  {\bibfield  {journal} {\bibinfo  {journal} {Nat. Nano.}\ }\textbf {\bibinfo
  {volume} {8}},\ \bibinfo {pages} {561--564} (\bibinfo {year}
  {2013})}\BibitemShut {NoStop}%
\bibitem [{\citenamefont {Lo~Nardo}\ \emph {et~al.}(2015)\citenamefont
  {Lo~Nardo}, \citenamefont {Wolfowicz}, \citenamefont {Simmons}, \citenamefont
  {Tyryshkin}, \citenamefont {Riemann}, \citenamefont {Abrosimov},
  \citenamefont {Becker}, \citenamefont {Pohl}, \citenamefont {Steger},
  \citenamefont {Lyon}, \citenamefont {Thewalt},\ and\ \citenamefont
  {Morton}}]{lonardo_2015}%
  \BibitemOpen
  \bibfield  {author} {\bibinfo {author} {\bibfnamefont {Roberto}\ \bibnamefont
  {Lo~Nardo}}, \bibinfo {author} {\bibfnamefont {Gary}\ \bibnamefont
  {Wolfowicz}}, \bibinfo {author} {\bibfnamefont {Stephanie}\ \bibnamefont
  {Simmons}}, \bibinfo {author} {\bibfnamefont {Alexei~M.}\ \bibnamefont
  {Tyryshkin}}, \bibinfo {author} {\bibfnamefont {Helge}\ \bibnamefont
  {Riemann}}, \bibinfo {author} {\bibfnamefont {Nikolai~V.}\ \bibnamefont
  {Abrosimov}}, \bibinfo {author} {\bibfnamefont {Peter}\ \bibnamefont
  {Becker}}, \bibinfo {author} {\bibfnamefont {Hans-Joachim}\ \bibnamefont
  {Pohl}}, \bibinfo {author} {\bibfnamefont {Michael}\ \bibnamefont {Steger}},
  \bibinfo {author} {\bibfnamefont {Stephen~A.}\ \bibnamefont {Lyon}}, \bibinfo
  {author} {\bibfnamefont {Mike L.~W.}\ \bibnamefont {Thewalt}}, \ and\
  \bibinfo {author} {\bibfnamefont {John J.~L.}\ \bibnamefont {Morton}},\
  }\bibfield  {title} {\enquote {\bibinfo {title} {Spin relaxation and
  donor-acceptor recombination of {Se}${}^+$ in 28-silicon},}\ }\href {\doibase
  10.1103/PhysRevB.92.165201} {\bibfield  {journal} {\bibinfo  {journal} {Phys.
  Rev. B}\ }\textbf {\bibinfo {volume} {92}},\ \bibinfo {pages} {165201}
  (\bibinfo {year} {2015})}\BibitemShut {NoStop}%
\bibitem [{\citenamefont {Morse}\ \emph {et~al.}(2016)\citenamefont {Morse},
  \citenamefont {Abraham}, \citenamefont {Riemann}, \citenamefont {Abrosimov},
  \citenamefont {Becker}, \citenamefont {Pohl}, \citenamefont {Thewalt},\ and\
  \citenamefont {Simmons}}]{morse_2016}%
  \BibitemOpen
  \bibfield  {author} {\bibinfo {author} {\bibfnamefont {Kevin~J.}\
  \bibnamefont {Morse}}, \bibinfo {author} {\bibfnamefont {Rohan J.~S.}\
  \bibnamefont {Abraham}}, \bibinfo {author} {\bibfnamefont {Helge}\
  \bibnamefont {Riemann}}, \bibinfo {author} {\bibfnamefont {Nikolai~V.}\
  \bibnamefont {Abrosimov}}, \bibinfo {author} {\bibfnamefont {Peter}\
  \bibnamefont {Becker}}, \bibinfo {author} {\bibfnamefont {Hans-Joachim}\
  \bibnamefont {Pohl}}, \bibinfo {author} {\bibfnamefont {Michael L.~W.}\
  \bibnamefont {Thewalt}}, \ and\ \bibinfo {author} {\bibfnamefont {Stephanie}\
  \bibnamefont {Simmons}},\ }\bibfield  {title} {\enquote {\bibinfo {title} {A
  photonic platform for donor spin qubits in silicon},}\ }\href@noop {} {\
  (\bibinfo {year} {2016})},\ \Eprint {http://arxiv.org/abs/1606.03488
  [cond-mat, physics:quant-ph]} {arXiv:1606.03488 [cond-mat, physics:quant-ph]}
  \BibitemShut {NoStop}%
\end{thebibliography}%

\newpage



\section*{Supplementary material}

# Optimizing surface defects for atomic-scale electronics: Si dangling bonds

## Supplemental material

Peter Scherpelz[1,*] and Giulia Galli[1,2]

[1]*Institute for Molecular Engineering,*

*The University of Chicago, Chicago, IL, USA*

[2]*Materials Science Division, Argonne National Laboratory, Argonne, IL, USA*

(Dated: May 25, 2017)

In this supplemental material we include detailed notes on the calculations which were reported in the main text, as well as convergence studies, and results at the PBE level when only hybrid results were shown in the main text.

## NEUTRAL DB STATE CALCULATIONS

### Calculation details

Our basic calculations use the generalized gradient approximation functional developed by Perdew, Burke, and Ernzerhof (PBE) [1]. All calculations use Troullier-Martins norm-conserving pseudopotentials [2] generated by D. Ceresoli [3] and a wavefunction energy cutoff of 40 Ry. The vacuum between slabs is 8.9 Å, and a $2 \times 2 \times 1$ unshifted Monkhorst-Pack [4] grid is used to sample the Brillouin zone (unshifted Monkhorst-Pack grids are used throughout this work).

The dielectric-based hybrid calculations used a 40 Ry wavefunction energy cutoff, and an 80 Ry cutoff for the exact exchange (Fock) operator. Only the $\mathbf{\Gamma}$-point was used to sample the Brillouin zone. Finally, $G_0W_0$ calculations were based on DFT calculations using the PBE functional. These calculations used a 30 Ry wavefunction energy cutoff, the $\mathbf{\Gamma}$-point only, and $N_{\text{PDEP}} = 3072$ [5].

For this section the silicon surface was modeled as a symmetric slab, with one DB on each surface. The initial lattice was built using the minimum-energy lattice constant for bulk Si using the PBE functional, which was found to be $5.4653 \pm 0.0003$ Å. Approximate positions for the monohydride termination were put in by hand, the appropriate hydrogen atoms were removed to create DBs, and atom positions were randomly perturbed to break any artificial symmetries. The full slab was then relaxed until forces on the atoms were less than $0.013 \, \text{eV}/\text{Å}$. The crystal used for the hybrid calculation was not separately relaxed. A $6 \times 6$ ($4 \times 4$) horizontal supercell was used for 12 layers and fewer (more than 12 layers). Here we use the notation $x \times y$ to indicate $x$ atoms perpendicular to the dimer rows by $y$ atoms parallel to the dimer rows in a given layer.

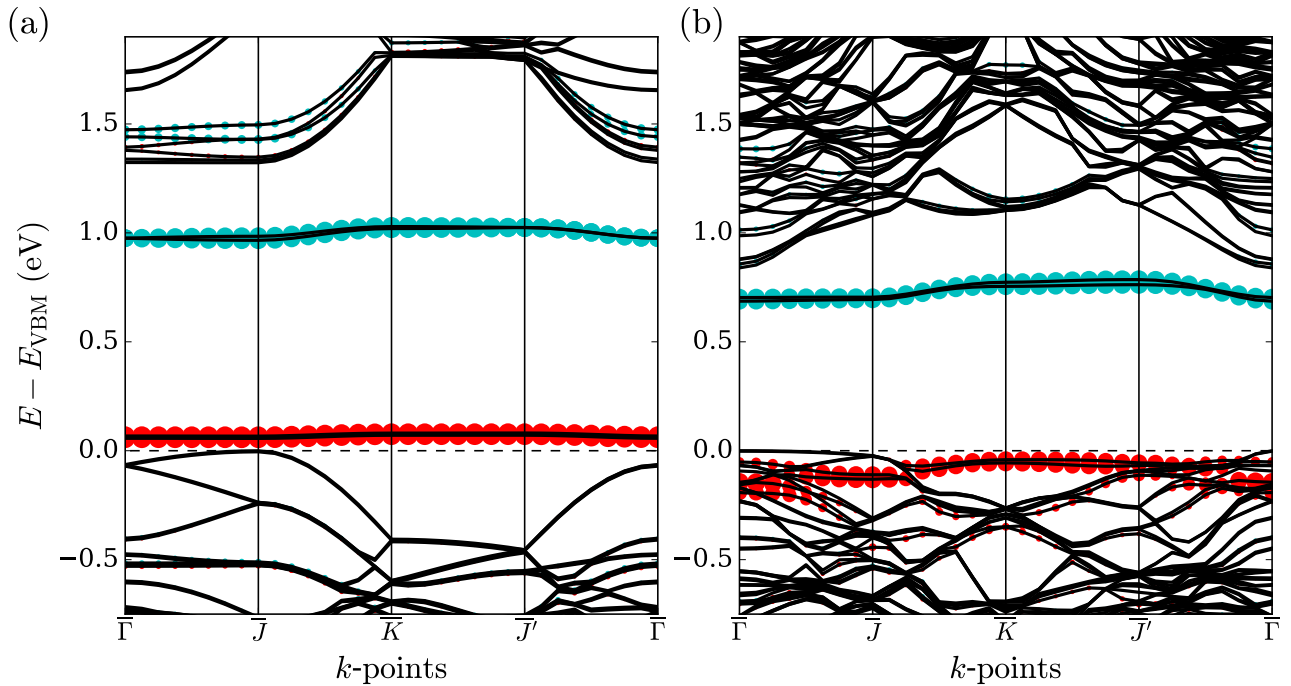

FIG. 1. Band plots corresponding to the **(a)** 4- and **(b)** 16-layer calculations shown in Figure 1(a) of the main manuscript. The circles show the proportion of overlap that a given Kohn-Sham eigenstate has with the atomic wavefunctions of the Si atoms with DBs; the circle area is proportional to the overlap, and red (cyan) colors correspond to the overlap for spin-up (spin-down) eigenstates.

**Identification of DB states**

Dangling bond states were identified by projecting the Kohn-Sham eigenstate wavefunction onto the orthogonalized wavefunctions of the Si atoms with DB states. Isolated DBs had overlaps $|\Psi|^2$ from 0.3 to 0.45. We chose a lower bound of 0.1 to identify a state as having some DB characteristics. While the exact threshold is arbitrary, in cases where this threshold identified multiple states with DB characteristics, rather than a single isolated DB state, it was visually clear that both states had some degree of mixing with bulk states, as in Figure 1(d) of the main manuscript.

In Figure 1(c) and 1(d) of the main manuscript, the isosurface identifies the surface at which $|\Psi|^2 = 0.0013/\text{Å}$ for the selected, occupied Kohn-Sham eigenstate, using DBs from the 8- and 16-layer results shown in Figure 1(a).

As an alternate way of visualizing DB character, Figure S1 shows the degree to which each band exhibits overlap with the atomic wavefunctions of the Si atoms with DBs. The

3

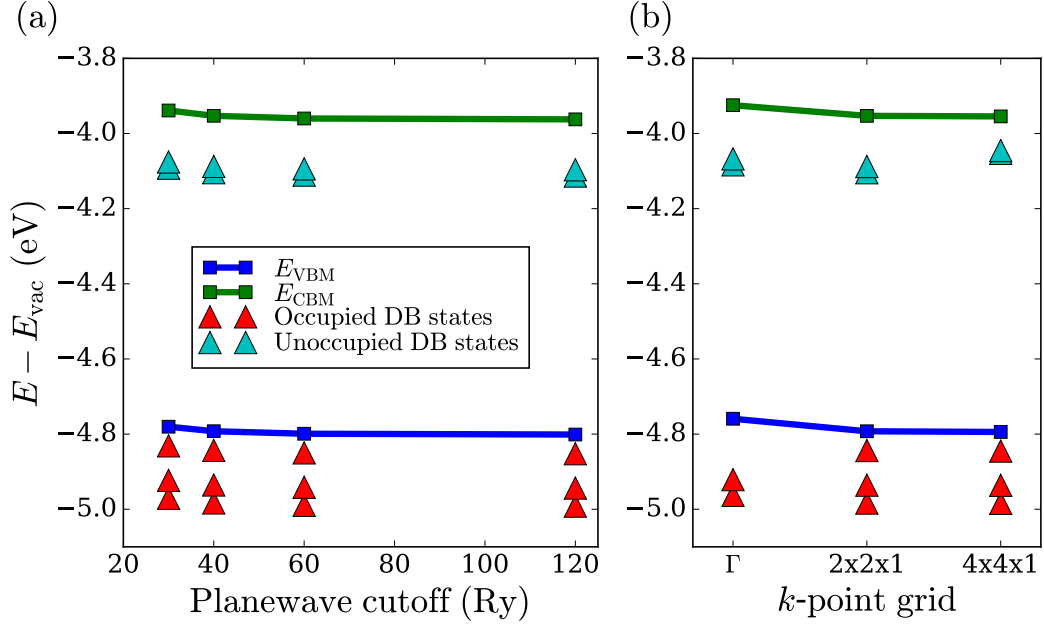

FIG. 2. Convergence of the dangling bond and band positions with regard to the vacuum energy, for 16-layer, $4 \times 4$ horizontal unit cell calculations, with symmetric DBs on both surfaces, as in Figure 1 of the main text. **(a)** Variation of the VBM, CBM, and DB energy levels with energy cutoff of the wavefunction. **(b)** Convergence with regard to the unshifted $k$-point mesh used. Note the "missing" occupied DB just below the VBM for the $\mathbf{\Gamma}$-point-only sampling is a result of a slightly lower overlap with atomic wavefunctions: The state has an overlap with atomic wavefunctions of 0.09, which puts it just below the 0.1 threshold used to plot DB states throughout this work. That state has $E = -4.82\,\text{eV}$, very close to the $E = -4.84\,\text{eV}$ value for the $4 \times 4 \times 1$ mesh.

4-layer plot clearly exhibits isolated DB states, while the 16-layer plot shows mixing of bulk and DB states, especially near the $\mathbf{\Gamma}$-point.

**Convergence studies**

Figure S2 shows the convergence with regard to the plane-wave energy cutoff for the wavefunctions, and the convergence with regard to the $k$-point mesh used. Overall, compared to 120 Ry, 30 Ry (40 Ry) calculations show deviations of 0.025 eV (0.010 eV). For $k$-points, $\mathbf{\Gamma}$-point-only shows deviations up to 0.035 eV, whereas $2 \times 2 \times 1$ values are within 0.002 eV, except for the unoccupied DB states, which seem to be more sensitive and differ by 0.05 eV.

4

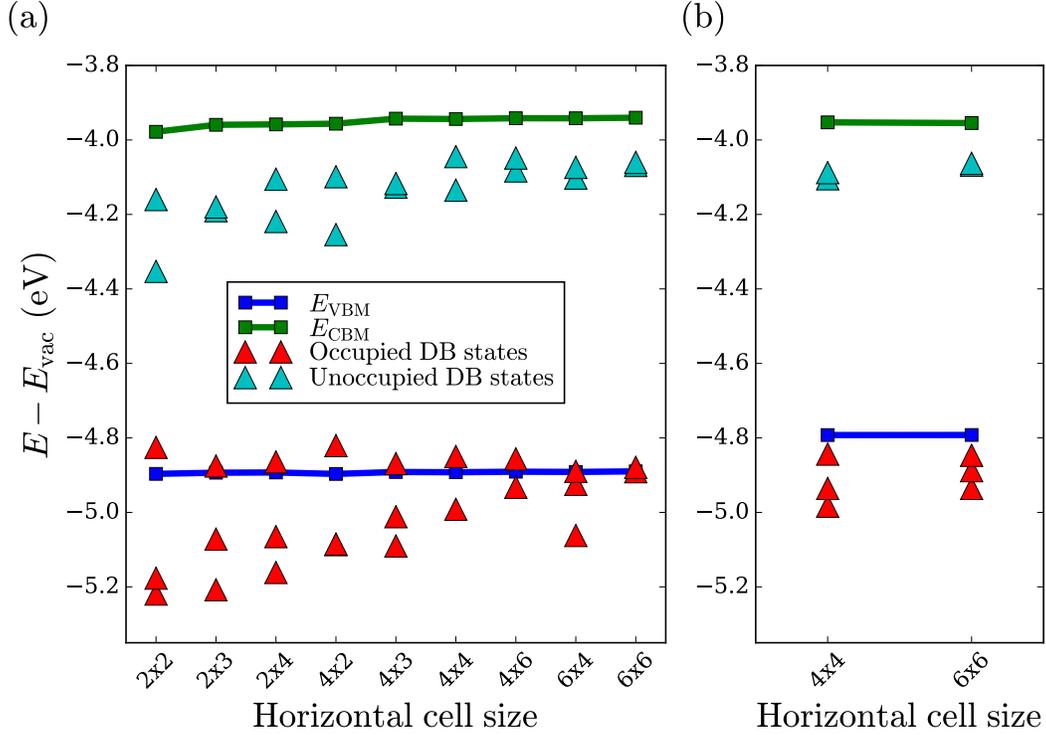

FIG. 3. Convergence of the dangling bond and band positions with regard to the vacuum energy, for symmetric DBs on both surfaces and other parameters as in Figure 1 of the main text. **(a)** Variation of the VBM, CBM, and DB energy levels for a 12-layer system **(b)** The same variation for two 16-layer systems.

Figure S3 shows convergence with regard to the horizontal supercell chosen. In all cases, a $4 \times 4$ supercell puts the VBM and CBM within 0.004 eV of the $6 \times 6$ supercell result. In general, the horizontal geometry primarily affects the degree of dispersion present in the DB states. In the 12-layer case shown in Figure S3(a), dispersion significantly affects DB positions near the VBM for supercells smaller than $6 \times 6$; in contrast, in Figure S3(b), DB positions are within 0.05 eV between $4 \times 4$ and $6 \times 6$ systems as the DB positions are no longer close to the VBM.

Other convergence parameters tested, but not plotted due to the minimal changes, are vacuum distance and the convergence threshold for crystal relaxation. For vacuum, we tested the 8.9 Å vacuum distance used in the paper against 17.8 Å; the VBM, CBM, and DB positions relative to vacuum all changed by 0.003 eV or less. For relaxation, we tested the 0.013 eV/Å used in this work against a threshold of 0.003 eV/Å, finding changes of 0.005 eV

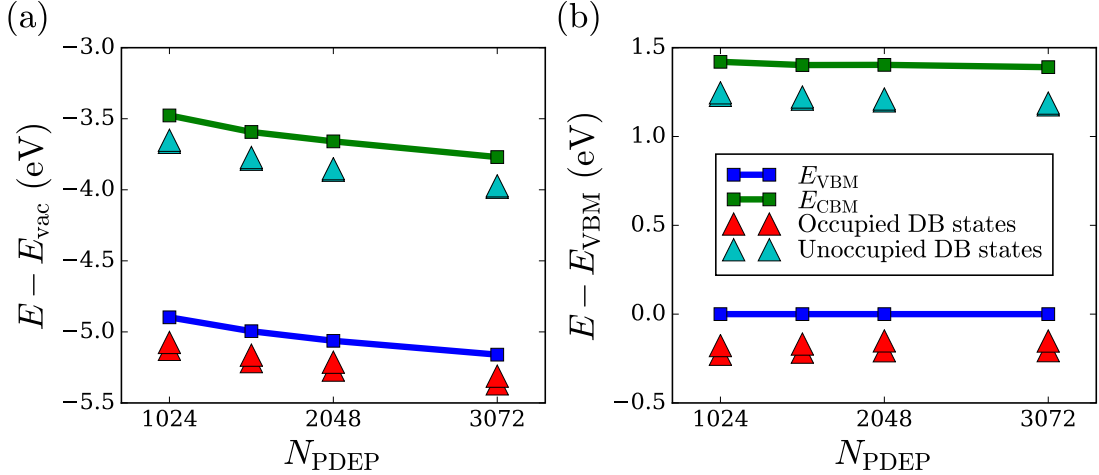

FIG. 4. Convergence of the dangling bond and band positions for varying $N_{\text{PDEP}}$, with other parameters as in Figure 1 of the main text. **(a)** Variation relative to the vacuum energy, showing changes of up to $0.12\,\text{eV}$ between $N_{\text{PDEP}} = 2048$ and $N_{\text{PDEP}} = 3072$. **(b)** Variation relative to the VBM energy, showing that the relative energy positions of all these states converge very quickly, with differences of $0.02\,\text{eV}$ or less between $N_{\text{PDEP}} = 2048$ and $N_{\text{PDEP}} = 3072$.

or less in the VBM, CBM, and DB positions relative to vacuum.

For the $G_0W_0$ calculation, convergence of $N_{\text{PDEP}}$ (see Ref. [5] for details of this parameter) is shown in Figure S4.

**CHARGE TRANSITION LEVELS**

For charge transition level calculations, a slab with bulk termination on one side, and a monohydride reconstruction with a single DB on the other side, was used. To create the bulk termination we started with a 34-layer slab, using a $2 \times 1$ horizontal supercell with a $4 \times 8 \times 1$ $k$-point mesh. This slab had all Si atoms in their bulk crystal positions. A dihydride hydrogen termination was then put in by hand, and relaxed to $0.013\,\text{eV}/\text{Å}$ with all Si atoms fixed.

Here one side of the slab was created with the bulk termination used above, and the remaining atoms were then separately relaxed for each charge state and layer number. In these relaxations, the hydrogen positions of the bulk termination were fixed, as were the bottom one, two, or three Si layers (for calculations of 4, 8, and 12 or more total Si layers,

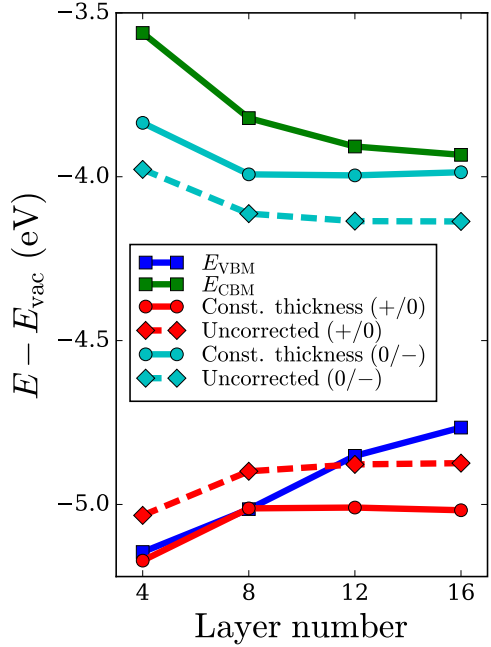

FIG. 5. Adiabatic charge transition levels calculated at the PBE level, otherwise matching the plot in Figure 2 of the main manuscript.

respectively); relaxation was performed for a $4 \times 4$ horizontal supercell with a $2 \times 2 \times 1$ $k$-point mesh, and a 40 Ry wavefunction cutoff, at the PBE level. For Figure 2 of the main paper, $4 \times 6$ horizontal supercells were used with a $\Gamma$-point only calculation, as $4 \times 4$ supercells were too small to align the electrostatic potentials far from the charged defect. A 30 Ry wavefunction cutoff was used, with a 60 Ry cutoff for the exact exchange calculation. The enlargement from $4 \times 4$ to $4 \times 6$ horizontal supercells was done by using the dimer positions furthest from the DB for the neutrally charged system for the extra two rows. The vacuum was enlarged to 16 Å for all of these calculations to mitigate the effects from charged surfaces.

The charge transition levels were corrected based on the method from Ref. [6]. We first applied a transverse electric field to the neutral system to determine the dielectric constant as a function of $z$. A model of this functional dependence was created following the procedure in Ref. [6]. A Gaussian was used for the charge density. The results were not sensitive to the width of this Gaussian, which was chosen as 0.6 Å (0.9 Å for negative charges at 12 and 16 layers, which were more spread out). Results were more sensitive to the $z$ position of the Gaussian charge distribution, so it was determined by minimizing the total absolute

difference between the planar-averaged model and DFT potentials (averaged over the $x$ and $y$ directions). Once this was determined, the potential $V$ was computed by solving the Poisson equation, and the electrostatic energy of the system calculated using $E = \frac{1}{2} \int d\mathbf{r}\rho(\mathbf{r})V(\mathbf{r})$ [6]. Finally, the energy correction was calculated as the difference between this energy for the periodic system, and the energy found by uniformly extrapolating the model system to infinite size [6]. Alignment of the electrostatic potential was performed by finding the difference between the potentials averaged over $x$ and $z$ at $y = y_\rho + L_y/2$, where $y_\rho$ is the position of the charge and $L_y$ the supercell size in this direction. Energies were referenced to the VBM of the neutral DB system, which differed from the VBM of the system without defects by $0.002$ eV or less.

Figure S5 shows the charge transition levels calculated at the PBE level, with parameters otherwise identical to that in Figure 2 of the main paper. The $(+/0)$ transition moves down in energy to some degree, but trends are very similar.

For convergence of this calculation, we checked other systems at the PBE level, including a $4 \times 8$ horizontal supercell with up to a $4 \times 2 \times 1$ $k$-point mesh, an $8 \times 8$ horizontal supercell for 4- and 8-layer systems, and a $4 \times 8$ supercell with $32$ Å of vacuum. While uncorrected charge transition levels varied by up to $0.3$ eV, corrected transition levels varied by $0.03$ eV to $0.06$ eV in most cases, with a maximum variation of $0.10$ eV for the 4-layer, $8 \times 8$ horizontal supercell.

## MULTIPLE DB CALCULATIONS

The calculation of a neutral DB pair and DB wire used the same bulk-like dihydride termination for the bottom of the slab as in the previous section, and the PBE functional. To assess convergence, we evaluated the position of the highest-lying occupied DB at the $\Gamma$-point relative to the VBM. For the DB pair, we used a $4 \times 6$ supercell, and a $2 \times 2 \times 1$ $k$-point mesh. These runs were compared to calculations using only the $\Gamma$ point for $k$-point sampling, finding $0.03$ eV variation in the position of the DB relative to the VBM at $\Gamma$. Convergence was also checked with $4 \times 10$ and $8 \times 6$ supercell geometries, for which a 30 Ry wavefunction cutoff and $\Gamma$-only $k$-point sampling was used. The changes in the DB position due to the choice of geometry were $0.012$ eV or less. A calculation using the dielectric-dependent hybrid functional was also performed, again using a 30 Ry wavefunction cutoff

8

and the Γ-point. Here DB positions varied by 0.06 eV or less.

For the DB wire, a $4 \times 2$ supercell was used with a $2 \times 4 \times 1$ $k$-point mesh. This was compared to a $4 \times 8 \times 1$ $k$-point mesh, finding 0.003 eV variation in the position of the DB relative to the VBM at Γ. Supercells of $4 \times 4$ ($2 \times 2 \times 1$ $k$-point mesh), $8 \times 2$ ($1 \times 4 \times 1$ $k$-point mesh), and (for the 4-layer system) $8 \times 4$ ($1 \times 2 \times 1$ $k$-point mesh) were performed, showing variation of 0.04 eV or less in the DB position when compared to the base calculation. Finally, calculations using the dielectric-dependent hybrid functional also showed variation of 0.04 eV or less in the DB position relative to the VBM.

The overlap was calculated and displayed as for Fig. S1, except that the coloring was the same for spin-up and spin-down eigenstates.

## STRAIN CALCULATIONS

For strained systems in the main manuscript, all calculations including relaxations were performed with the HSE hybrid functional. Bulk calculations were used to determine the minimum-energy lattice constant for HSE as $5.4364 \pm 0.0003$ Å. Biaxial strain was then applied to bulk crystals, and the $z$-direction lattice constant that produced a minimum energy was found. For each strain value, the appropriate $z$-direction lattice constant was found to an accuracy of between $\pm 0.0008$ Å (for 0.5% strain) and $\pm 0.005$ Å (for 3% strain).

Figure 4(a) in the manuscript used a slab that was bulk-terminated on both sides. This bulk termination was created by relaxing hydrogen atoms on an 18-layer, fixed Si slab for each strain value. Layers were then added or subtracted from the Si slab to generate 10-, 14-, and 34-layer slabs; no relaxation of Si atoms was performed. In these calculations, a $2 \times 1$ horizontal supercell was used with a $4 \times 8 \times 1$ $k$-point mesh. A 40 Ry wavefunction cutoff was used with an 80 Ry cutoff for the exact exchange calculation.

Figure 4(b) of the manuscript used a $4 \times 4$ horizontal supercell with Γ-point only sampling [7]. One surface had the typical monohydride reconstruction with a single DB; bulk termination was used on the other side. Parameters, including lattice constants and hydrogen positions for the bulk termination, otherwise matched those in Figure 4(a). A relaxation of the crystal was performed using the HSE hybrid functional, with the bottom H and Si layers fixed as for the bulk termination in Figure 2.

PBE calculations were also performed for the calculations in Figure 4(a), starting with

9

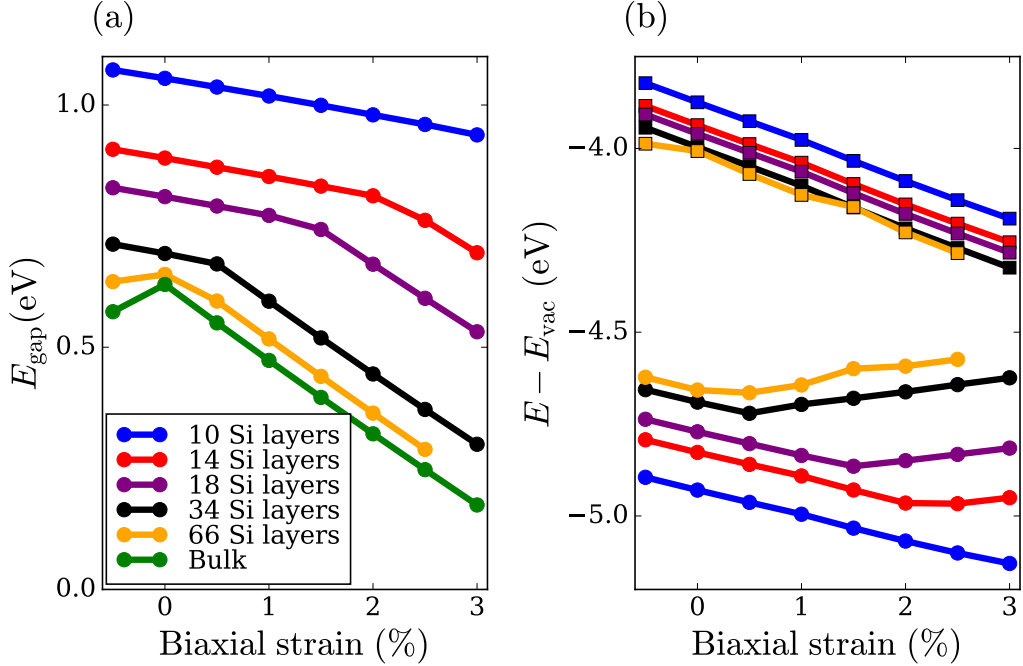

FIG. 6. Variation of energy levels with strain, based on PBE-level calculations; other parameters match those of Figure 4 of the manuscript. **(a)** Variation of the band gap vs. strain. **(b)** Variation of the VBM (circles) and CBM (squares) positions vs. strain.

the minimum-energy lattice constant for PBE, and using the bulk-termination hydrogen positions calculated for Figure 2 of the manuscript. These results are shown in Figure S6. The minimum VBM positions relative to vacuum are the same as for the HSE results.

For these thin slabs, the choice of termination does seem to have an effect on the biaxial strain at which the minimum VBM energy is found. This is shown for slabs created with monohydride termination on both sides at the PBE level in Figure S7. The atomic positions were relaxed with no atoms fixed in this case. While the minimum VBM energy moves to lower strain values, the qualitative behavior is similar. The most physically relevant case for a silicon-on-insulator system is likely that of one monohydride surface, i.e. that in Figure 4(b) of the main manuscript, which shows a VBM minimum at 2.5% tensile strain for a 10-layer slab.

---

* pscherpelz@uchicago.edu

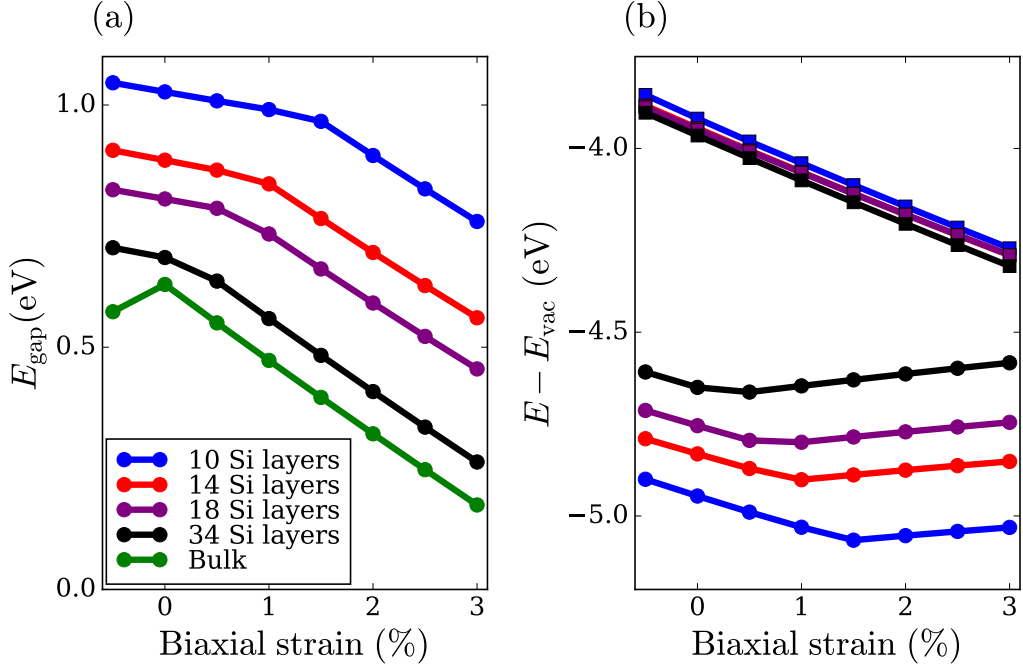

FIG. 7. Variation of energy levels with strain, based on PBE-level calculations, for slabs terminated with monohydride reconstructions and fully relaxed. **(a)** Variation of the band gap vs. strain. **(b)** Variation of the VBM (circles) and CBM (squares) positions vs. strain.

[1] John P. Perdew, Kieron Burke, and Matthias Ernzerhof, "Generalized gradient approximation made simple," Phys. Rev. Lett. **77**, 3865–3868 (1996).

[2] N. Troullier and José Luís Martins, "Efficient pseudopotentials for plane-wave calculations," Phys. Rev. B **43**, 1993–2006 (1991).

[3] Davide Ceresoli, "Pseudopotentials," https://sites.google.com/site/dceresoli/pseudopotentials, (accessed May 20, 2016).

[4] Hendrik J. Monkhorst and James D. Pack, "Special points for brillouin-zone integrations," Phys. Rev. B **13**, 5188–5192 (1976).

[5] Marco Govoni and Giulia Galli, "Large scale GW calculations," J. Chem. Theory Comput. **11**, 2680–2696 (2015).

[6] Hannu-Pekka Komsa and Alfredo Pasquarello, "Finite-size supercell correction for charged defects at surfaces and interfaces," Phys. Rev. Lett. **110**, 095505 (2013).

[7] For biaxial strains of 1.5%, 2.0%, and 2.5%, technical convergence problems led us to use a $2 \times 2 \times 1$ $k$-point mesh, but with only a single **q** vector, $\mathbf{q} = 0$.



\end{document}